\documentclass{emulateapj}
\usepackage{natbib}

\newcommand\etal{et al.}
\newcommand\ie{i.e.}
\newcommand\eg{e.g.}
\newcommand\msun{{\rm M_\odot}}
\newcommand\kms{\ifmmode{\rm km\ s^{-1}}\else$\rm km\ s^{-1}$\fi}
\newcommand\mydeg{$^\circ$}
\def\eps@scaling{1.0}
\newcommand\plotsix[6]{{
 \typeout{Plotfour included the files #1 #2 #3 #4 #5}
 \centering
 \leavevmode
 \columnwidth=.3\columnwidth
 \includegraphics[width={\eps@scaling\columnwidth}]{#1}
 \includegraphics[width={\eps@scaling\columnwidth}]{#2}
 \includegraphics[width={\eps@scaling\columnwidth}]{#3}
 \includegraphics[width={\eps@scaling\columnwidth}]{#4}
 \hspace{20pt}
 \includegraphics[width={\eps@scaling\columnwidth}]{#5}
 \hspace{20pt}
 \includegraphics[width={\eps@scaling\columnwidth}]{#6}
}}

\shorttitle{RADIO AND X-RAY STUDY OF A133}
\shortauthors{RANDALL ET AL.}
\slugcomment{Accepted for publication in the Astrophysical Journal}

\begin{document}

\title{Radio and Deep {\it Chandra} Observations of the Disturbed Cool
Core Cluster Abell~133}

\author{S.\ W.\ Randall\altaffilmark{1}, T.\ E.\ Clarke\altaffilmark{2}, P.\ E.\ J.\ Nulsen\altaffilmark{1}, M.\ S.\ Owers\altaffilmark{3}, C.\ L.\ Sarazin\altaffilmark{4}, W.\ R.\ Forman\altaffilmark{1}, S.\ S.\ Murray\altaffilmark{1}}

\begin{abstract}
We present results based on new {\it Chandra} and multi-frequency
radio observations of the disturbed cool core cluster Abell~133.  The
diffuse gas has a complex bird-like morphology, with a plume of
emission extending from two symmetric wing-like features.  The plume is
capped with a filamentary radio structure that has been previously
classified as a radio relic.  
X-ray spectral fits in the region of the relic indicate the presence
of either high temperature gas or non-thermal emission, although the
measured photon index is flatter than would be expected if the
non-thermal emission is from IC scattering of the CMB by the radio
emitting particles. 
We find evidence for a weak elliptical
X-ray surface brightness edge surrounding the core, which we show is
consistent with a sloshing cold front.  The plume is consistent with
having formed due to uplift by a buoyantly rising radio bubble, now
seen as the radio relic, and has properties consistent with buoyantly
lifted plumes seen in other systems (e.g., M87).  Alternatively, the
plume may be a gas sloshing spiral viewed edge on.  Results from
spectral analysis of the wing-like features are inconsistent with the
previous suggestion that the wings formed due to the passage of a weak
shock through the cool core.  We instead conclude that the wings are
due to X-ray cavities formed by displacement of X-ray gas by the radio relic.
The central cD galaxy contains two small-scale cold gas clumps that
are slightly offset from their optical and UV counterparts, suggestive
of a galaxy-galaxy merger event.  On larger scales, there is evidence
for cluster substructure in both optical observations and the X-ray
temperature map.  We suggest that the Abell~133 cluster has recently
undergone a merger event with an interloping subgroup, initiating gas
sloshing in the core.  The torus of sloshed gas is seen close to
edge-on, leading to the somewhat ragged appearance of the elliptical surface
brightness edge.
We show that the additional buoyant force from a passing subcluster
can have a significant effect on the rise trajectories of buoyant
bubbles, although this effect alone cannot fully explain the morphology
of Abell~133.
The radio observations reveal a large-scale double-lobed structure not previously
identified in the literature.  We conclude that this structure
represents a previously unreported background giant radio galaxy at 
$z = 0.293$, the 
northern lobe of which overlies the radio relic in the core of
Abell~133.  A rough estimate indicates that the contribution of this
background lobe to 
the total radio emission in the region of the relic is modest ($< 13$\%).

\end{abstract}

\altaffiltext{1}{Harvard-Smithsonian Center for Astrophysics, 60
  Garden St., Cambridge, MA 02138, USA; srandall@cfa.harvard.edu}
\altaffiltext{2}{Naval Research Laboratory, 4555 Overlook Ave. SW, Code 7213, Washington, DC 20375, USA}
\altaffiltext{3}{Centre for Astrophysics and Supercomputing, Swinburne University, P.O. Box 218, Hawthorn, VIC 3122, Australia}
\altaffiltext{4}{Department of Astronomy, University of Virginia, P.O. Box 3818, Charlottesville, VA 22903-0818}

\section{Introduction} \label{sec:intro}

Abell~133 is an X-ray luminous
cluster at $z = 0.0566$ (Struble \& Rood 1999) with indications of a
central cooling flow (White \etal\ 1997).  
X-ray emission from Abell~133 was first detected by the {\it Ariel 5}
satellite (Cooke \etal\ 1977).
In the radio, the cD contains a point-like source associated with some
extended emission, and a filamentary diffuse structure roughly 40~kpc
northwest of the cD with an unusually steep spectral index, which has been
classified as a radio relic (Slee \etal\ 2001, hereafter S01).  
{\it ROSAT} and {\it Very Large Array} ({\it VLA}) observations revealed a
correlation between central bright X-ray and diffuse radio emission
(Rizza \etal\ 2000).  Rizza \etal\ (2000) detect a bridge of
1.4~GHz radio emission connecting the central source and the diffuse
relic, although S01 find no evidence for a connection based on
higher fidelity observations at the same frequency.

More recently, {\it
  Chandra} (Fujita \etal\ 2002, hereafter F02) and {\it XMM-Newton} (Fujita \etal\
2004, hereafter F04) observations revealed a complex
bird-like morphology in the central diffuse gas, with a plume of
emission extending from symmetric wing-like features at the core.  The plume
is capped by the diffuse radio structure, and there is a clear deficit
of X-ray emission in the region of the relic (excluding the
region of the overlapping plume).
F02 and F04 consider several formation mechanisms for the plume and
wings.  They conclude that the plume was likely formed due to
lifting of cool central gas by the buoyantly rising radio bubble,
similar to what is seen in M87.  With the addition of {\it XMM-Newton}
observations, F04 suggest that the wings are formed by the
passage of a weak shock through the cool core.  As the shock front passes
through the dense central gas, it curves around the central density
peak, giving rise to the edges that define the X-ray wings.  They also
place an upper limit on non-thermal emission from inverse Compton (IC)
scattering of cosmic microwave background (CMB) photons by
relativistic radio-emitting particles in the region of the relic.

In this paper, we report on results from a new {\it Chandra}
observation of Abell~133 and from multi-frequency {\it VLA} radio observations.
Based on these new observations, we re-interpret the origin of
the structure in the diffuse gas, examine the connection between the X-ray
and radio features, and draw conclusions about the dynamical state and
history of Abell~133.
We assume an angular diameter distance to Abell~133 of 223.7~Mpc, which
gives a scale of 
1.08~kpc/\arcsec .
All error ranges are 68\% confidence intervals (\ie, 1$\sigma$), unless
otherwise stated.

\section{Observations and Data Reduction} \label{sec:obs}

\subsection{X-ray Observations} \label{sec:x-ray}

Abell~133 was originally observed with {\it Chandra} on October 13,
2000, for 36~ksec with the {\it Chandra} CCD Imaging Spectrometer
(ACIS), pointed such that the central core was visible on the
back-side illuminated ACIS-S3 CCD.  It was subsequently observed twice
in June 2002 with the front-side illuminated
ACIS-I CCD array for a total of 90~ksec, with 
the pointing offset such the core was just beyond the corner of the field of
view (FOV), to study the extended diffuse cluster emission
(Vikhlinin \etal\ 2005).  Most recently, Abell~133 was observed on
August 29, 2008, for 70~ksec with the ACIS-I CCD array, pointed such that the
cluster core was well within the FOV.  We consider the data from all
four of these observations whenever possible, although the core itself is
only covered by the ACIS-S3 and most recent ACIS-I observations.
The details of these observations are given in Table~\ref{tab:xray}.
These data were reduced using the method we have applied previously,
\eg, in Randall \etal\ (2008).  All data were reprocessed from the
level 1 event files using the latest 
calibration files (as of {\sc CIAO4.1}). CTI and time-dependent
gain corrections were applied where applicable. {\sc LC\_CLEAN} was
used to remove background
flares\footnote{\url{http://asc.harvard.edu/contrib/maxim/acisbg/}}.
The mean event rate was calculated using time bins within 3$\sigma$ of the
overall mean, and bins outside a factor of 1.2 of this mean were
discarded. The resulting cleaned exposure times were 30~ksec for
ACIS-S3, 43~ksec and 42~ksec for the ACIS-I offset pointings, and
69~ksec for the targeted ACIS-I observation, giving total cleaned
exposure times of 99~ksec in the core and 184~ksec west of the core.

Diffuse emission from Abell~133 fills the 
image FOV for each observation.  We therefore used the
standard {\sc CALDB\footnote{\url{http://cxc.harvard.edu/caldb/}}}
blank sky background files appropriate for each observation,
normalized to our observations in the 10-12 keV energy band.  To
generate exposure maps, we assumed a MEKAL model with $kT = 4$~keV,
Galactic absorption, and abundance of 30\% solar at a redshift $z =
0.0566$, which is consistent with typical results for the extended
emission from detailed
spectral fits (see \S~\ref{sec:xspec}).

\subsection{Radio Observations} \label{sec:radio}

We obtained low frequency radio observations of Abell~133 (PI: Clarke)
using the {\it VLA}. Observations at frequencies of 74 and 330~MHz were
obtained in the extended A configuration 
on August 16, 2003, 
and 330~MHz observations in B
configuration were taken on July 9, 2002. Details of the observations
are presented in Table~\ref{tab:radio}.

The low frequency observations were all taken in multi-channel
continuum mode to allow excision of radio frequency interference and
reduce the effects of bandwidth smearing. The A configuration 74~MHz
data used Cygnus~A as a bandpass and gain calibrator, while the 330~MHz
A configuration observations were calibrated using 3C48 as a bandpass,
phase, and flux calibrator. The B configuration 330~MHz observations
used 3C48 as the bandpass and flux calibrator and the nearby
calibrator source 0116-208 as a phase calibrator. All data were
processed using the NRAO Astronomical Image Processing System
(AIPS). The images were produced through the standard Fourier
transform deconvolution method. The data were processed through
several loops of imaging and self-calibration to reduce the effects of
phase and amplitude errors. Due to the large fields of view at these
low frequencies all data were processed within AIPS using the wide-field
imaging techniques which correct for distortions in the image caused
by the non-coplanarity of the {\it VLA}. This technique involved using a set
of overlapping maps (facets) to cover the desired image area (Cornwell
\& Perley 1992).

The first round of phase calibration of the 74 MHz data was undertaken
using the best 330 MHz B configuration images as the input model. This
method corrects any phase offsets between the two data sets. All data
sets were checked for positional errors introduced during calibration
by comparing positions from Gaussian fits of ten compact sources
surrounding the target to known source positions from the
literature. The 330 MHz A configuration data was found to have a small
position shift which was corrected using the known source positions.

In addition to our low frequency observations of this system, we have
also extracted and reduced {\it VLA} archive data at 1400 MHz taken in C
configuration. The data were recorded and analyzed in multi-channel
continuum mode. The observations included 3C48 as a flux and bandpass
calibrator and 0116-208 as a phase calibrator. As with the low
frequency observations, data were processed within AIPS following the
standard techniques including several rounds of phase-only, followed
by amplitude and phase self-calibration. Additionally, we include
higher resolution 1400 MHz radio data from Slee \etal\ (2001)
in our analysis.

\section{Image Analysis} \label{sec:imgs}

\subsection{The X-ray Image} \label{sec:ximg}

\begin{figure}
\plottwo{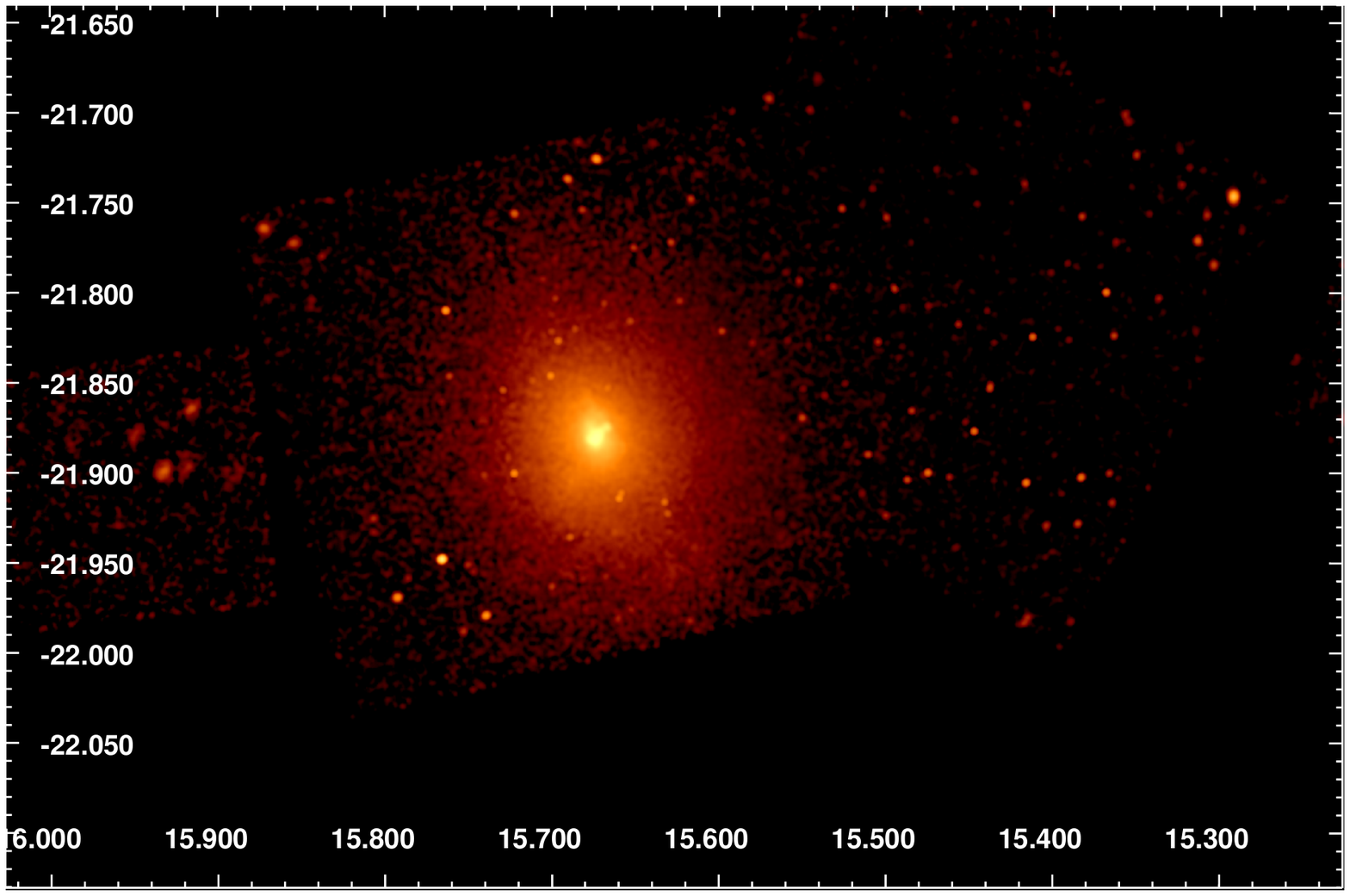}{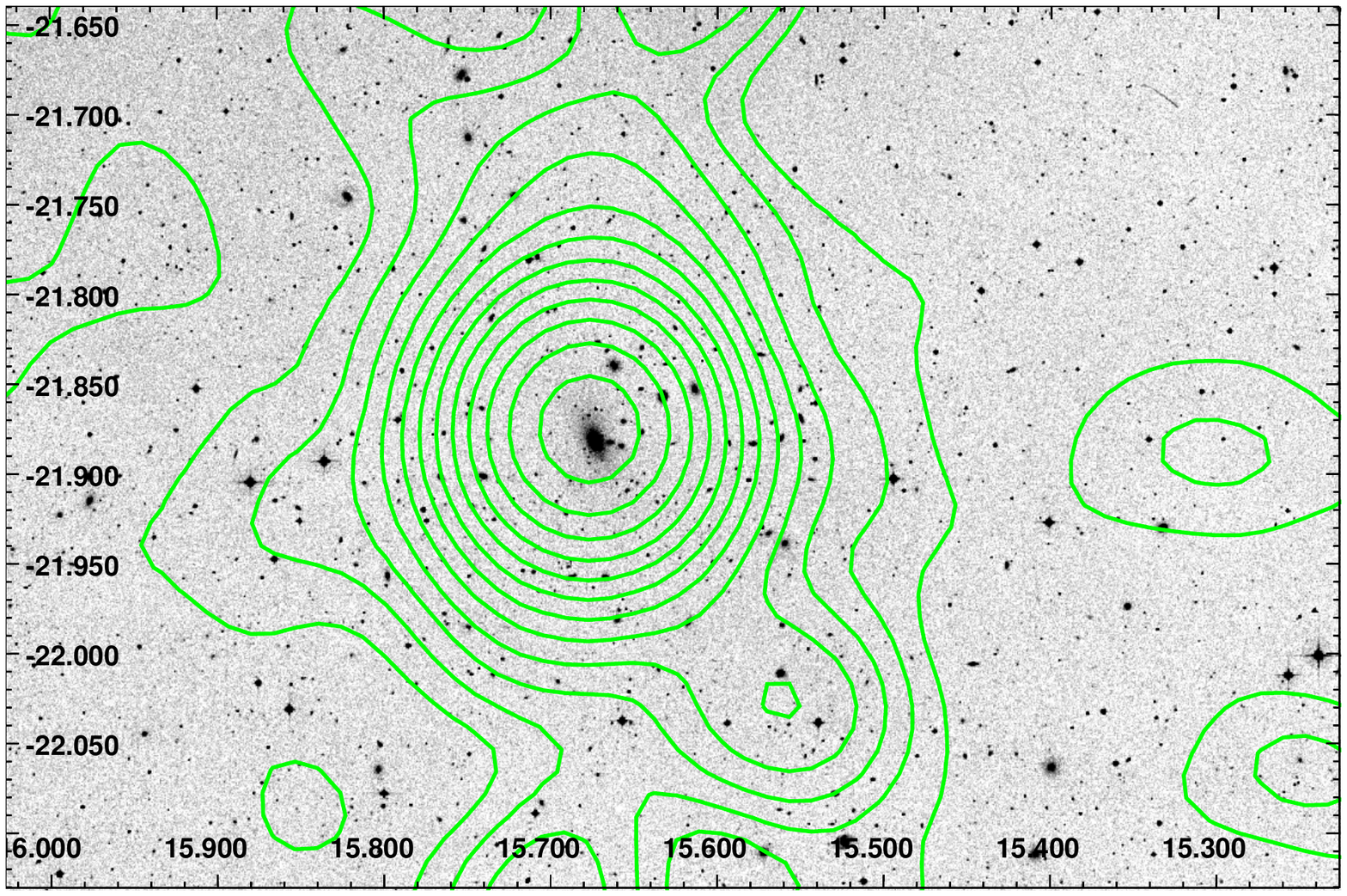}
\caption{
  {\it Left Panel:} Exposure corrected, background subtracted 0.6--5
  keV {\it Chandra} mosaic of Abell~133.  The image has
  been smoothed with an 9\arcsec\ radius Gaussian to better show the
  faint extended emission.    Regions with less than 
  10\% of the total exposure were omitted.
  The image shows complex structure in the core, and an elliptical
  morphology to the extended diffuse emission, elongated to the
  northeast and southwest.
  {\it Right Panel:} {\it DSS} image of the same field, with {\it
    RASS} contours overlaid.  The cD galaxy is seen near the centroid
  of the {\it RASS} contours.  The X-ray contours show a secondary
  peak 10.66\arcmin\ (694~kpc) southwest of the cD, beyond the {\it
    Chandra} field of view.
  \label{fig:fullimg}
}
\end{figure}

\begin{figure}
\plotone{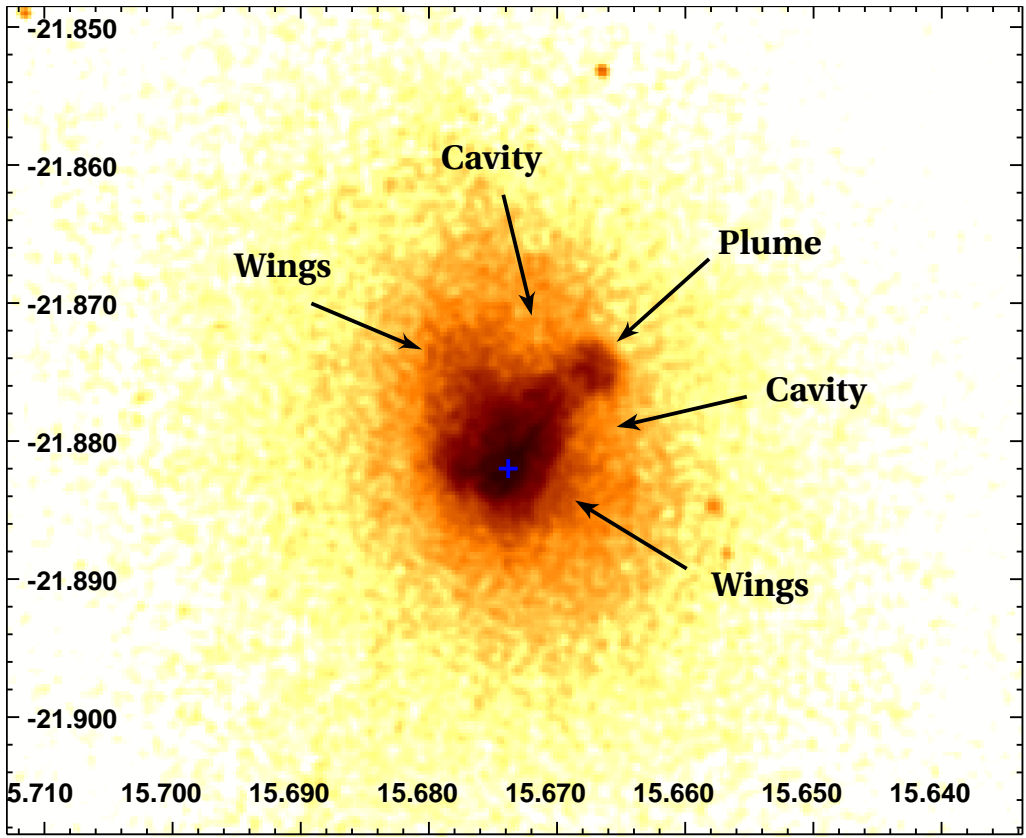}
\caption{
  Central region of the same image shown in Figure~\ref{fig:fullimg}
  ({\it Left}), smoothed with a 2\arcsec\ Gaussian.  The core shows a
  complex ``bird-like'' morphology, with wings to the northeast and
  southwest, a long plume of emission to the northwest, and 
  surface brightness depressions (cavities) between the wings and the
  plume.  The
  blue cross marks the position of the cD galaxy.
  \label{fig:xr_core}
}
\end{figure}

The exposure corrected, background subtracted, smoothed X-ray image is shown
in the left panel of Figure~\ref{fig:fullimg}.  In the right panel,
the optical {\it DSS} 
image of the same field is shown with {\it ROSAT All Sky
  Survey} ({\it RASS}) contours overlaid, which show the extended
emission from Abell~133 outside of the {\it Chandra} FOV.  The image
shows complex structure in the core, and an ellipticity in the
extended cluster emission, elongated to the north-northeast and south-southwest,
that is not evident in the {\it ROSAT} contours (although they do show
a second peak to the southwest, along the elongation axis).
A close-up of the
X-ray emission from the core is shown in Figure~\ref{fig:xr_core}.
The core shows a complex ``bird-like'' morphology, previously noted by
F02, with diffuse wings of emission extending from the southern end of
a long plume oriented from southeast to northwest (the
``tongue'' identified by F02).  On either side of the plume, north of
the X-ray wings, there are two surface brightness depressions, which
were identified as X-ray cavities (presumably devoid of X-ray emitting
gas) by B\^{i}rzan \etal\ (2004; 2008).  Although the cD galaxy
(marked with a blue cross in Figure~\ref{fig:xr_core}, position from
{\it 2MASS} catalog by Skrutskie \etal\ 2006) is near the
peak of diffuse X-ray emission, the overall emission is not
centered on the galaxy.  There is a finger of bright emission
extending from the brightness peak in the southeast along the interior
of the plume,
and terminating just before the round cap to the plume, giving the
impression that the plume has formed from gas being pushed or
pulled out of the central core.  The complex morphology of the diffuse
emission, as well as the offset of the central cD galaxy from the
local X-ray centroid, indicate a highly disturbed, non-relaxed core.

\begin{figure}
\plotone{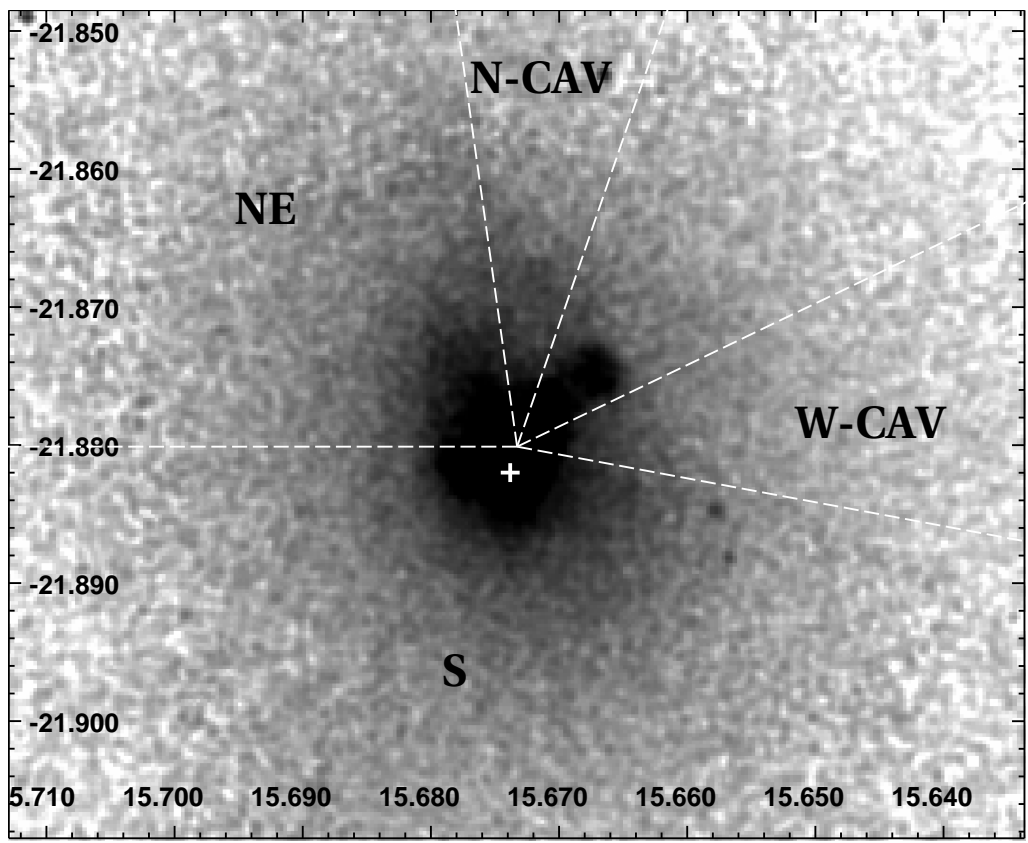}
\caption{
  X-ray image of the core with the sectors used for surface brightness
  profile extraction overlaid (dashed white lines).  The sector
  containing the plume was excluded.  The white cross marks the
  position of the cD galaxy.
  \label{fig:sectors}
}
\end{figure}

\begin{figure}
\plottwo{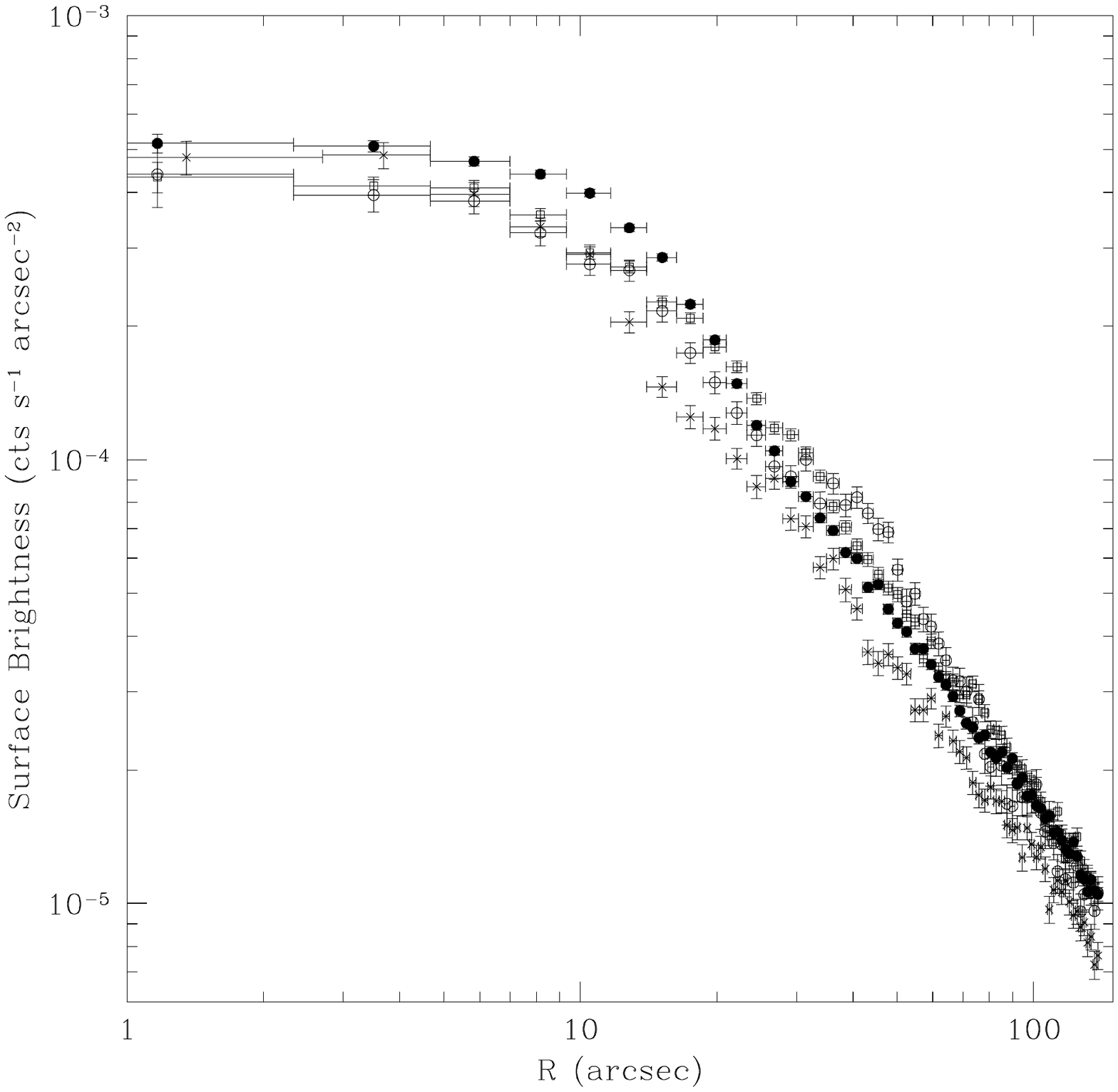}{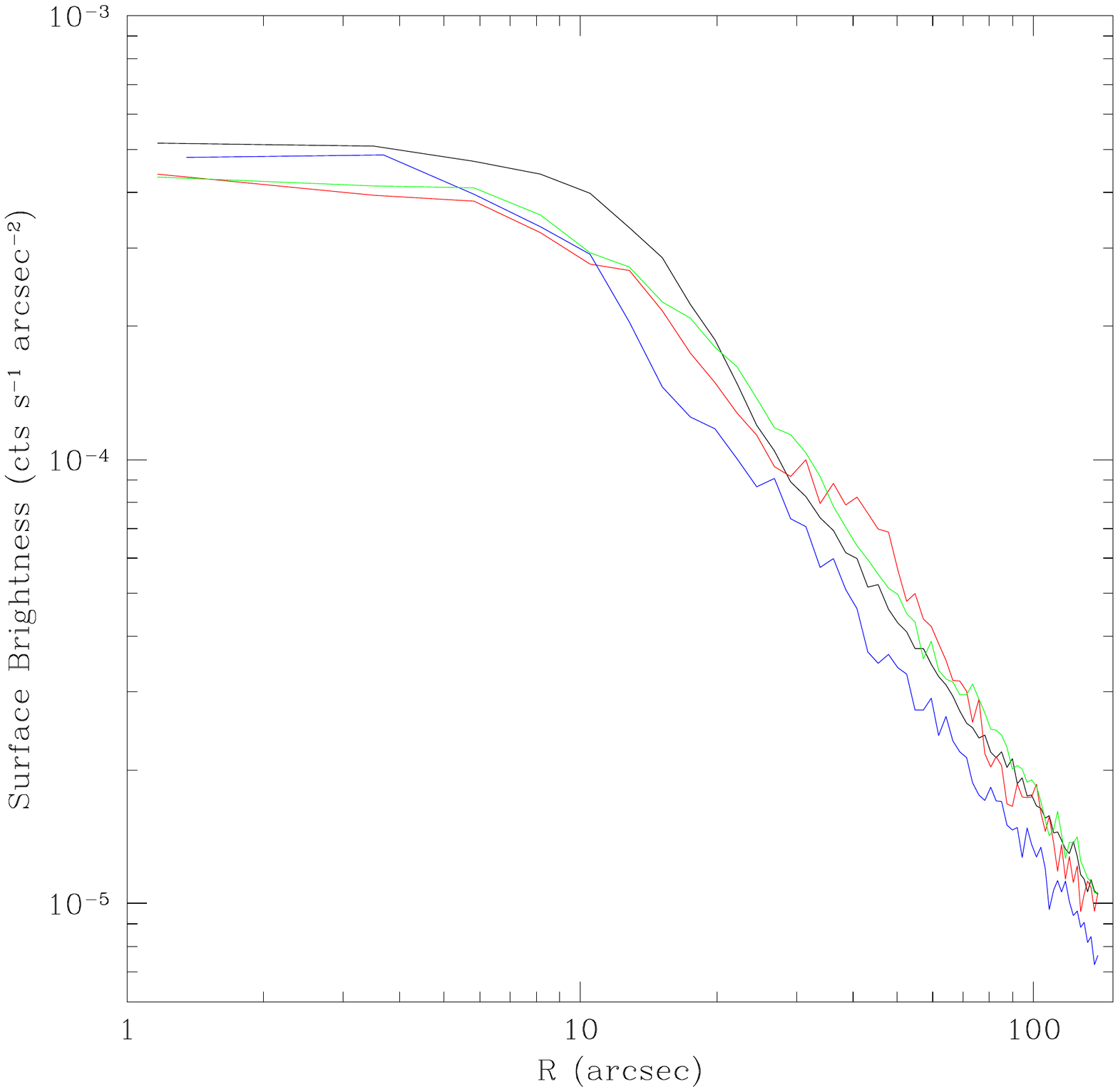}
\caption{
  Exposure corrected, background subtracted 0.6--5 {\it Chandra}
  surface brightness profile in the four sectors defined in
  \S~\ref{sec:ximg}: W-CAV (crosses, blue line), N-CAV (open 
  circles, red line), NE (open squares, green line), S (filled circles,
  black line)
  {\it Left Panel:} Profiles with error bars and radial ranges shown.
  {\it Right Panel:} Profiles shown as lines without error bars, for clarity.
  \label{fig:sbsect}
}
\end{figure}

To verify the surface brightness depressions on
either side of the plume, we extracted the
radial surface brightness profiles in four sectors: one for each of
the depressions, and the remaining azimuthal range split into two
sectors to the northeast and south (the sector containing the plume was
excluded). These sectors are shown in Figure~\ref{fig:sectors}.
The complex structure in the diffuse emission makes it difficult to
chose a central position for our extraction regions.  We choose to
use the centroid of the X-ray emission within 30\arcsec\ (33~kpc) 
of the core, which we find to be (01$^{\rm h}$02$^{\rm m}$41.6$^{\rm
  s}$, -21\mydeg52\arcmin48.4\arcsec\ [J2000]), 9\arcsec\ from the
position adopted by F02 based on the centroid of the emission at
smaller radii.  As can be seen from Figure~\ref{fig:fullimg}, this
position does not coincide with the central cD galaxy, nor does it
coincide with any bright peaks in the X-ray emission, but rather it
is offset from the cD galaxy in the direction of the plume.  However,
it is consistent with the centroid of the extended emission at large
radii, and is therefore appropriate for studying surface
brightness profiles outside of the core.  
The 0.6--5~keV exposure corrected surface
brightness profiles for concentric annuli in each sector are shown in
Figure~\ref{fig:sbsect}.  The left panel shows the measurements, with
error bars, while the right panel shows the profiles as connected lines,
without error bars, for clarity.  Each of the sectors corresponding to
surface brightness depressions show deficits between roughly
10-30\arcsec\ (red and blue lines in the right panel), consistent with
the X-ray image and with existing size
estimates for the X-ray cavities (B\^{i}rzan \etal\ 2004) and
coincident radio lobes (B\^{i}rzan \etal\ 2008) given in the
literature.  Such depressions are not seen in the northern and
southern sectors (green and black lines in the right panel of
Figure~\ref{fig:sbsect}), which instead show abrupt changes in slope at
about 27\arcsec.  We discuss these surface brightness
discontinuities further in \S~\ref{sec:profs}.

To quantify the significance of the cavities, we measured the net
surface brightness in annular sectors across the cavities
(corresponding to the sectors shown in Figure~\ref{fig:sectors}), and
in a 247\mydeg\ wide sector excluding the cavities and bright plume.
The radial range of the sectors was 10\arcsec -- 30\arcsec.  We find
average 0.3--2~keV net rates of $1.071 \pm 0.027  \times 10^{-4}$~cts~s$^{-1}$~pix$^{-1}$
for the western cavity,  $1.369 \pm 0.033 \times
10^{-4}$~cts~s$^{-1}$~pix$^{-1}$ for the northern cavity, and $1.660
\pm 0.010\times 10^{-4}$~cts~s$^{-1}$~pix$^{-1}$ for the wide sector.
Thus, the decrement associated with the northern cavity is significant
at 8.3$\sigma$, while the western cavity is significant at 20$\sigma$.

\subsection{Radio Images} \label{sec:rimg}

\begin{figure}
\plotone{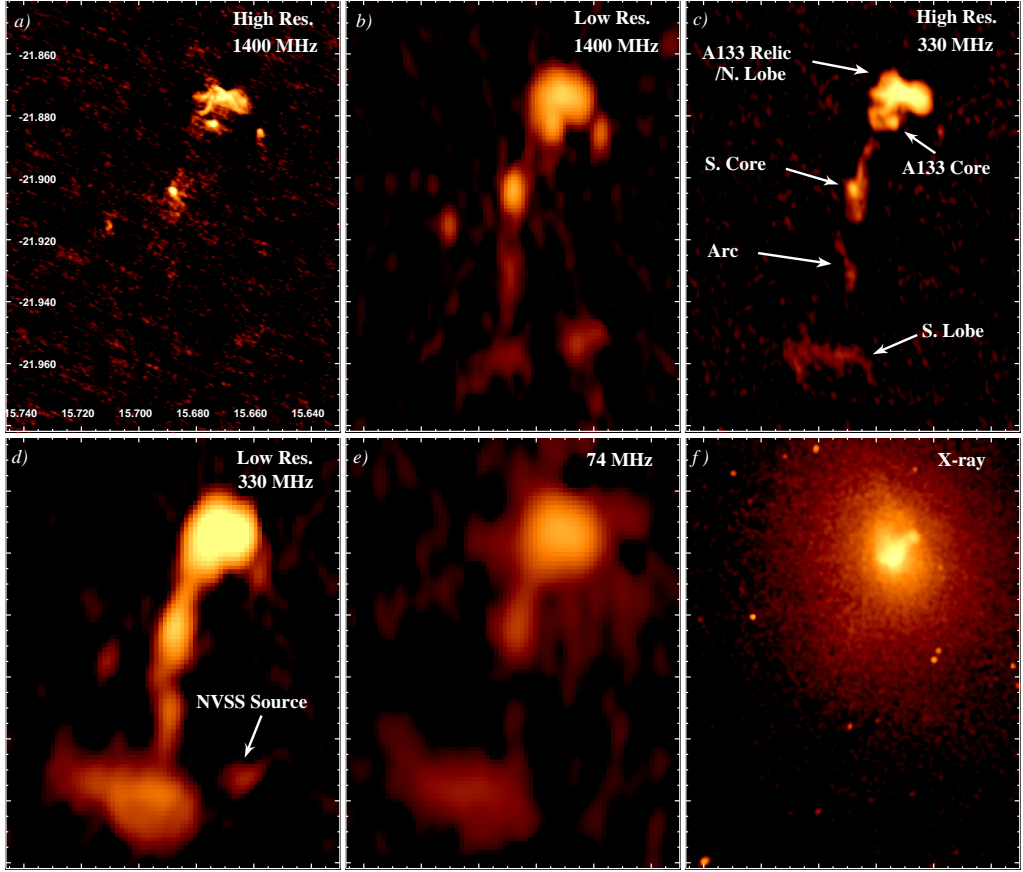}
\caption{
  {\it VLA} Radio images of Abell~133, at multiple frequencies and array
  configurations. {\it a)} 1400~MHz B configuration L-band,
  from S01. {\it b)} 1400~MHz C configuration
  L-band.  {\it c)} 330~MHz A configuration P-band. {\it d)} 
  330~MHz B configuration P-band. {\it e)} 74~MHz A
  configuration 4-band. {\it f)} 0.6--5.0~keV X-ray image,
  smoothed with a 4\arcsec\ Gaussian, for comparison.
  \label{fig:rad_imgs}
}
\end{figure}

Radio images derived from {\it VLA} observations at multiple frequencies and
array configurations are shown in Figure~\ref{fig:rad_imgs}.
The high-resolution 1.4~GHz image in Figure~\ref{fig:rad_imgs}a (taken from
S01) clearly shows the well-known diffuse radio source NVSS~J010241-215230 just
north of the core in Abell~133 (throughout this paper we will refer to
this source as the ``relic'', or sometimes the ``northern lobe'',
depending on context).  The relic exhibits complex
filamentary structure.  The core 
source in Abell~133 is roughly 34\arcsec\ (37~kpc) 
southeast of the relic.  The position of the core matches the optical
position of the cD galaxy.  There is some extended emission associated
with this core, in particular to the east, where there is a bright
inner 6.7\arcsec\ (7.2~kpc) extension connecting to a fainter loop
feature that extends 24\arcsec\ (26~kpc) from the core.
Based on older 1400~MHz {\it VLA} observations
(which we do not consider here), Rizza \etal\ (2000) claim to detect a
bridge of emission connecting the relic with the cD galaxy.  S01 argue
that this claim is not supported by their more 
sensitive, higher-resolution {\it VLA} observations (shown in
Figure~\ref{fig:rad_imgs}a).  Our 330~MHz A configuration
image shown in Figure~\ref{fig:rad_imgs}c does appear to show a
connection between
the cD and the relic, although the resolution is lower than in Figure~\ref{fig:rad_imgs}a.
Figure~\ref{fig:rad_imgs}a 
also shows two other prominent core-like sources, one 54\arcsec\
(59~kpc) west of the cD (which we we call the ``western core''), with
some extended emission to the south, and 
a second 1.53\arcmin\ (100~kpc) southeast of the cD (which we will
call the ``southern core''), with some
extended emission to the southwest.  Both sources were noted by S01
(galaxies E and J, respectively, in their Figure~6), and, 
as they point out, optical identifications put the 
western source within the cluster ($z = 0.0563$, Merrifield \& Kent
2001), while southeastern source is at a much higher redshift
($z=0.293$, Owen \etal\ 1995).

The lower frequency observations (and the lower-resolution 1400~MHz
image shown in Figure~\ref{fig:rad_imgs}b) reveal
unexpected structure in the large-scale radio emission.
The southern core shows two opposite extensions to the north
and south (which we will call the ``jets'' of the southern core), with
the northern jet
extending to the relic in Abell~133 in Figures~\ref{fig:rad_imgs}d,e.
A diffuse southern lobe is revealed, roughly 3.2\arcmin\
south of the southern core, somewhat larger than the distance to
the relic in the north (2.2\arcmin).  The lobe has a
shell-like morphology in Figures~\ref{fig:rad_imgs}b,c (low-resolution 1400~MHz
and high-resolution 330~MHz images), but appears more uniform
in Figures~\ref{fig:rad_imgs}d,e (low-resolution 330~MHz and 74~MHz images).
It is visible in Figure~10 of B\^{i}rzan \etal\ 2008, although they do
not discuss this feature.
It is distinct from the source reported in the NRAO {\it VLA}
Sky Survey ({\it NVSS}, Condon \etal\ 1998), NVSS~J010239-215708, indicated in
Figure~\ref{fig:rad_imgs}d and
also evident in Figure~\ref{fig:rad_imgs}b (we note that the {\it
  NVSS} source is
coincident with the position of the optically detected cluster
APMCC~138 reported by Dalton
\etal\ [1997] at $z = 0.06$, close to the redshift of Abell~133).
The southern jet of the southern core points in the direction of the
southern lobe, and connects to the lobe in
Figures~\ref{fig:rad_imgs}b,d.  The jet shows multiple twists, turns,
and bright spots, most evident in Figures~\ref{fig:rad_imgs}c,d (the
``arc'' indicated in Figure~\ref{fig:rad_imgs}c).

\begin{figure}
\plotone{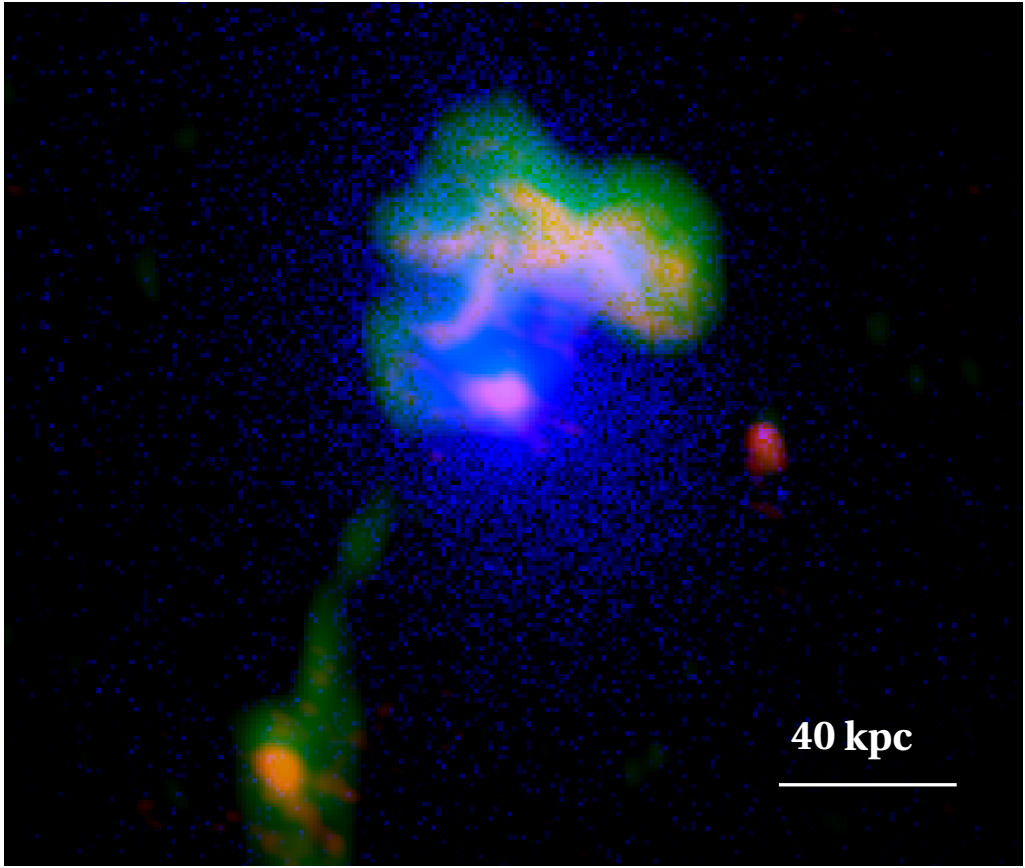}
\caption{
  Overlay of the {\it Chandra} X-ray (blue), 1400~MHz B configuration
  (red), and 330~MHz A configuration (green) images of the core of
  Abell~133.  The X-ray plume is capped with radio emission, which
  fills the X-ray cavities on either side of the plume.  The low
  frequency observations reveal a bridge between the southern core and
  the relic near the core of Abell~133.
  \label{fig:rad_overlay}
}
\end{figure}

The large-scale morphology of the low frequency radio observations is
consistent with what is expected for a double-lobed radio galaxy.  If
this is the case, the redshift of the optical source associated with
the southern core ($z=0.293$) gives a total projected extent of
1.6~Mpc (6.2\arcmin), measured from the outer edge of the southern
lobe to the outer edge of the northern lobe/relic.
However, as pointed out by F02, the relic feature to the north is
obviously associated with the X-ray core and plume in Abell~133.
This is clearly seen in Figure~\ref{fig:rad_overlay}, which  shows the
{\it Chandra} X-ray image (blue)
overlaid with the 1400~MHz B configuration image (red) and 330~MHz A
configuration (green).  The plume is capped by radio emission, which
overlaps the cavities discussed in \S~\ref{sec:ximg}, and is brightest in the
northern cavity.  The bridge of emission between the high-redshift
southern core and Abell~133 is also evident.  
The
low frequency radio observations therefore reveal that the radio emission
in the region of Abell~133 is a complicated combination of possibly
overlapping foreground and background sources.  We discuss these
features further in \S~\ref{sec:lobes}.

\section{X-ray Spectra} \label{sec:xspec}

The X-ray image (Figure~\ref{fig:fullimg}) shows complicated structure
in the core of Abell~133, as well as fainter ICM emission filling
the FOV.  We generated a
temperature map as a guide for detailed spectral fitting to
disentangle the various components and study the structure seen in the
ICM.  Spectra are grouped to a minimum of 40 counts per bin throughout.

\subsection{Temperature Map} \label{sec:tmap}

\begin{figure}
\plotone{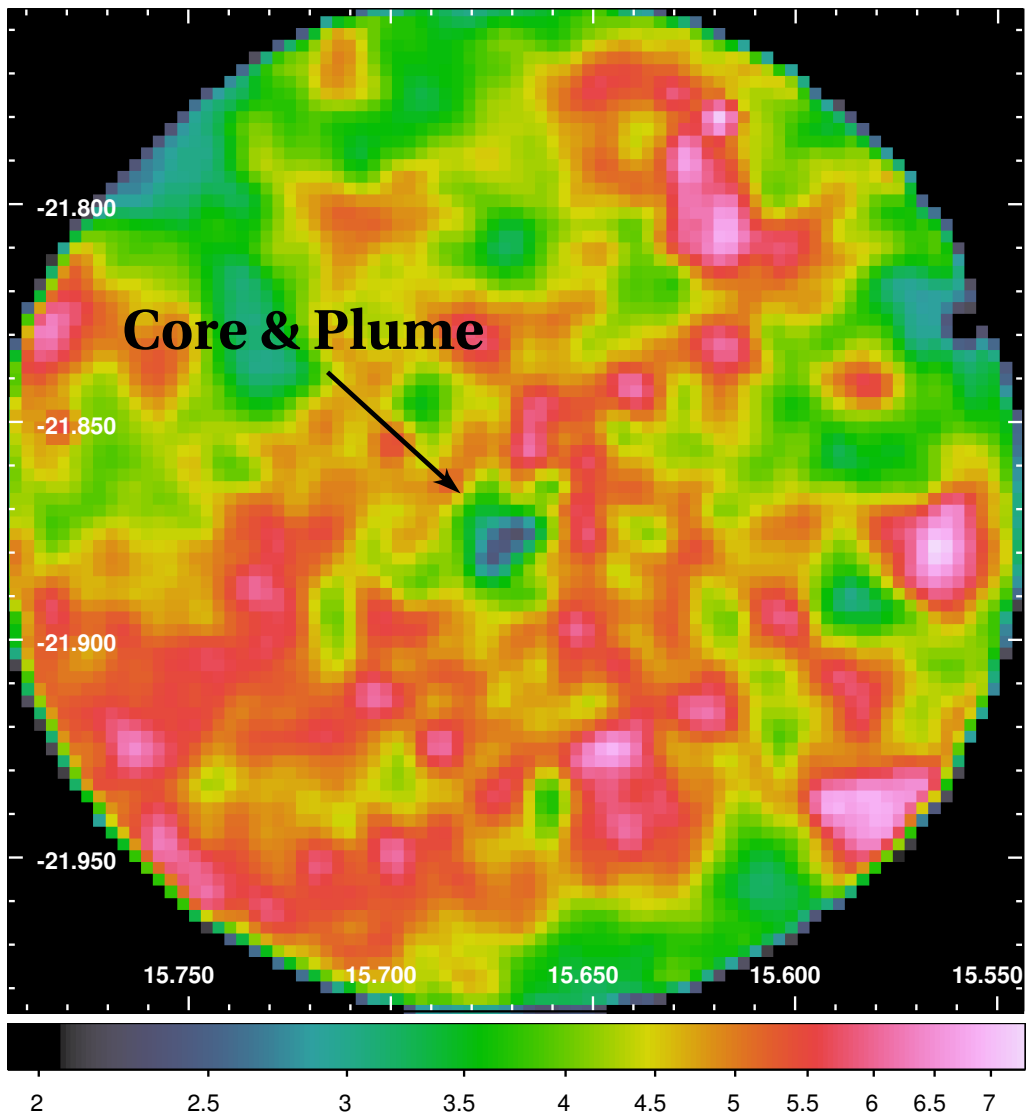}
\caption{Temperature map derived from {\it Chandra} X-ray
  observations, smoothed with a 2~pixel Gaussian. The scale is
  9.8~arcsec$^2$~pix$^{-1}$. 
  The color-bar gives the temperature in keV.
\label{fig:tmap}
}
\end{figure}

The temperature map was derived using the same method as developed in
Randall \etal\ (2008).  For each temperature map pixel, we
extracted a spectrum from a circular region containing 1000 net counts
(after subtracting the blank sky background).  The resulting spectrum
was fit in the 0.6 -- 7.0~keV range with an absorbed {\sc apec} model using
{\sc xspec}, with the abundance allowed to vary.  Our detailed
spectral fits indicate an absorbing column of about 1.8 times the
Galactic value of $N_H = 1.57\times 10^{20}$~cm$^{-2}$, consistent with
what was found by F02.  We therefore fixed the hydrogen column density
at $N_H = 2.85\times 10^{20}$~cm$^{-2}$, which we find from detailed
fits to the diffuse emission within the central 1.86\arcmin\ (121~kpc)
presented in \S~\ref{sec:dxspec}. The 
resulting temperature map is shown in Figure~\ref{fig:tmap}.    Faint
regions had extraction radii 
on the order of 1.4\arcmin\ (91~kpc), while the brightest regions,
near the core, had radii of 5\arcsec\ (5.4~kpc).  Even with the coarse
binning in the temperature map (9.8~arcsec$^2$~pix$^{-1}$) the
``bird-like'' morphology of the central core can be seen, with cool
wings of emission to the northeast and southwest, on either side of
the cooler plume extending to the northwest.  There is
a curling tail of cooler emission extending east of the core, which
is not associated with the radio bridge to the southern core noted
in \S~\ref{sec:rimg}.  Overall, the diffuse emission is patchy, with
regions of cooler and hotter gas with statistically significant
temperature differences.  There is little large-scale structure in the
map, expect perhaps on average higher temperatures in the
southern half of the map as compared to the north.

\begin{figure}
\plotsix{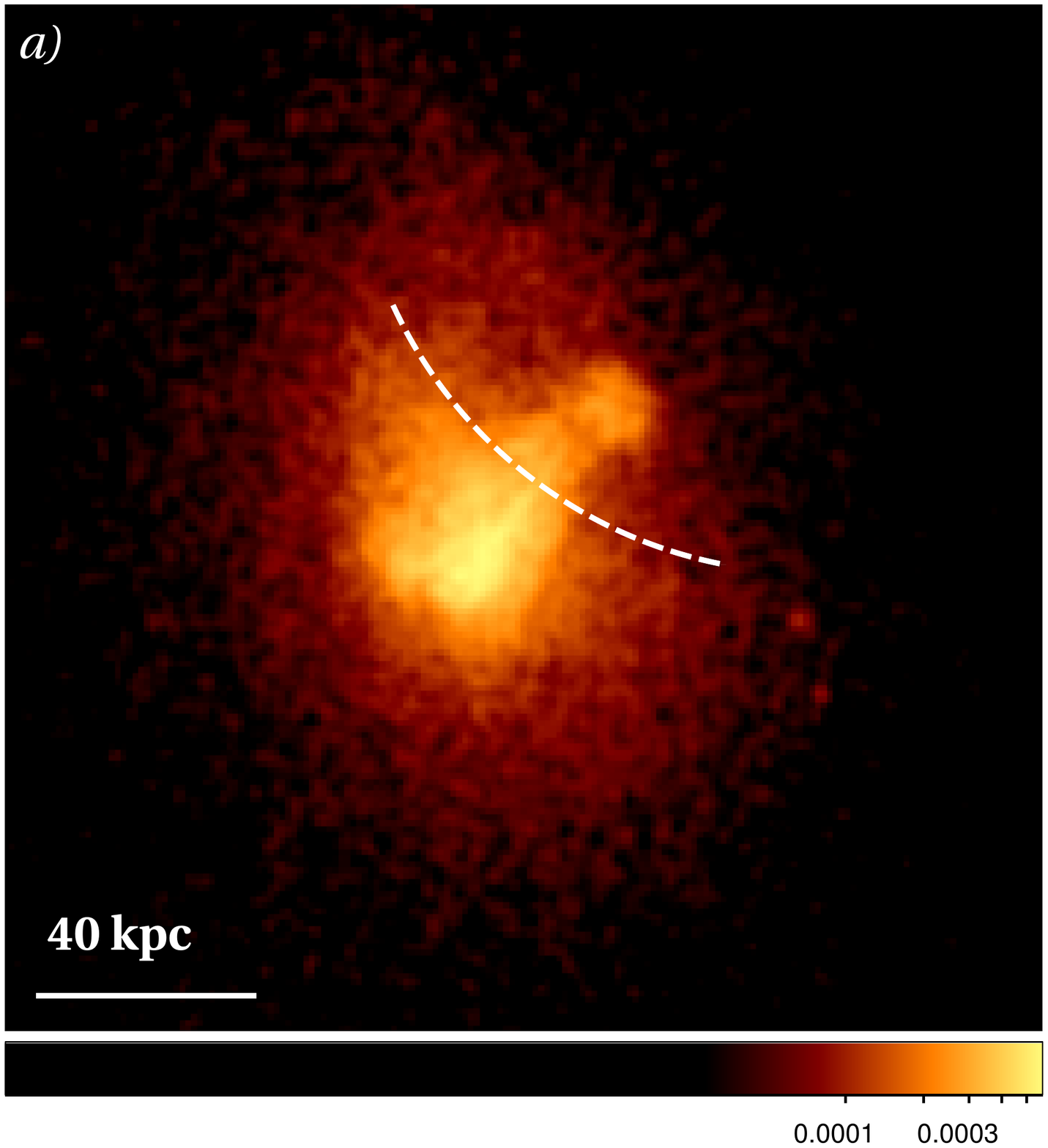}{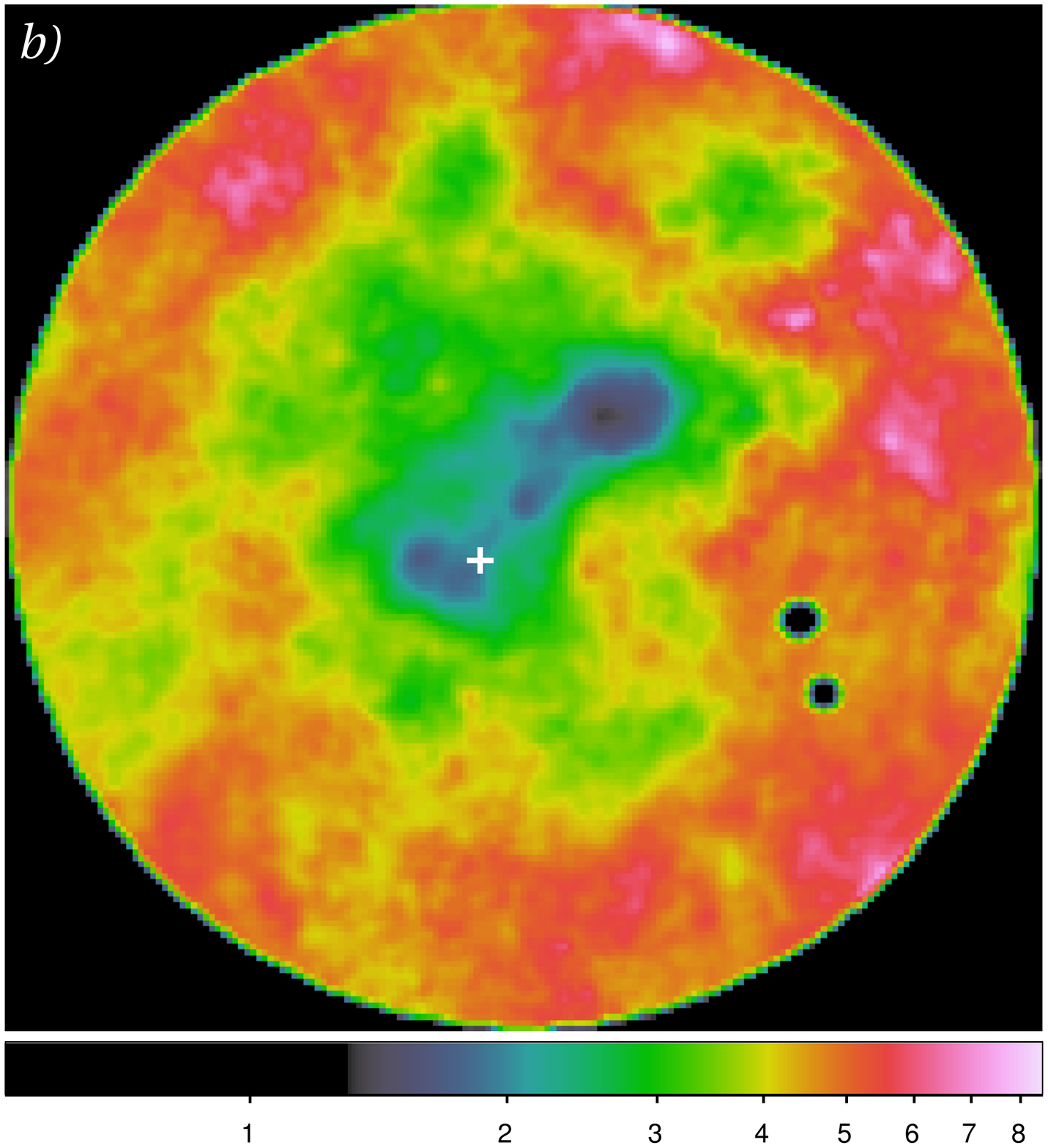}{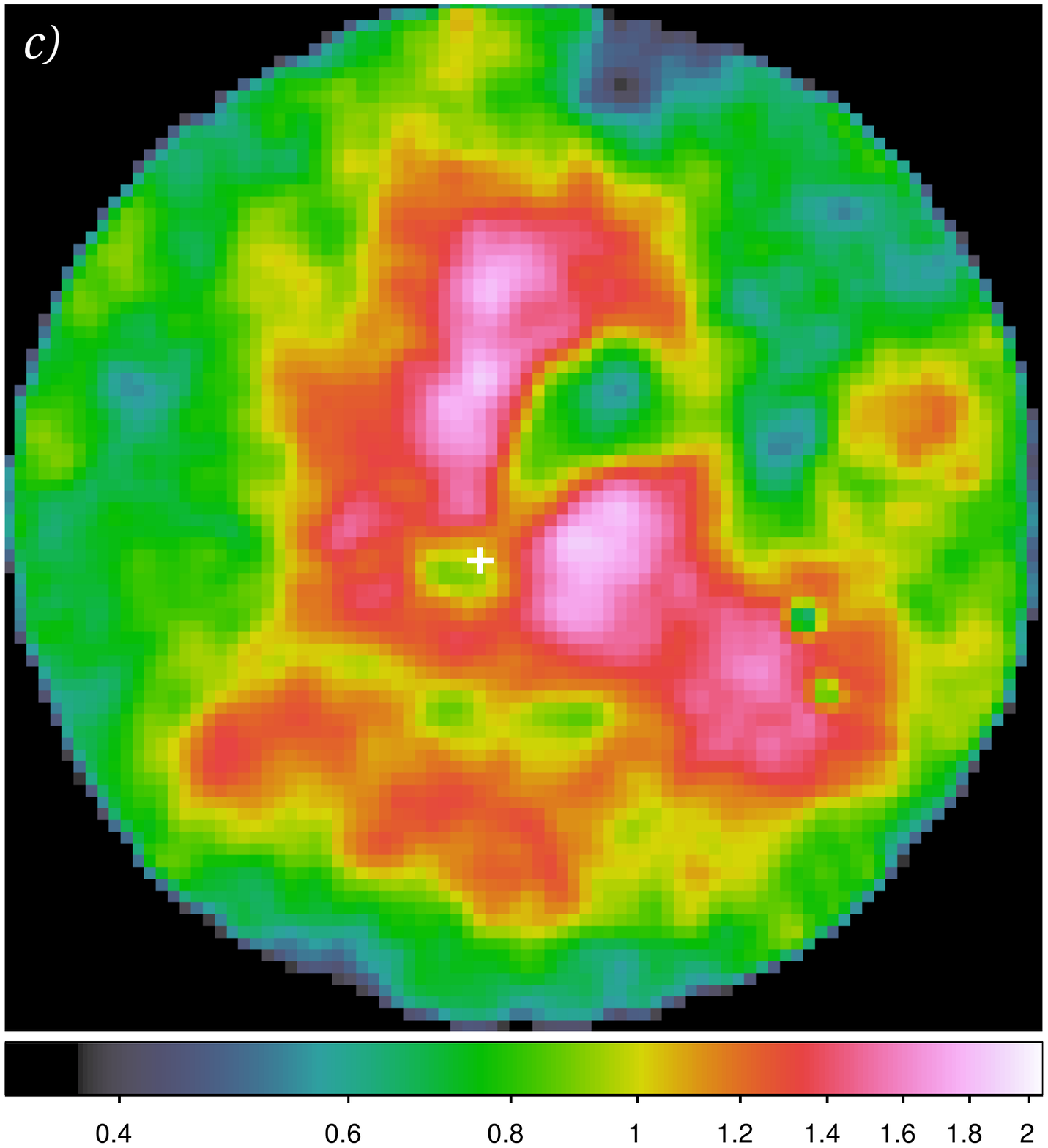}{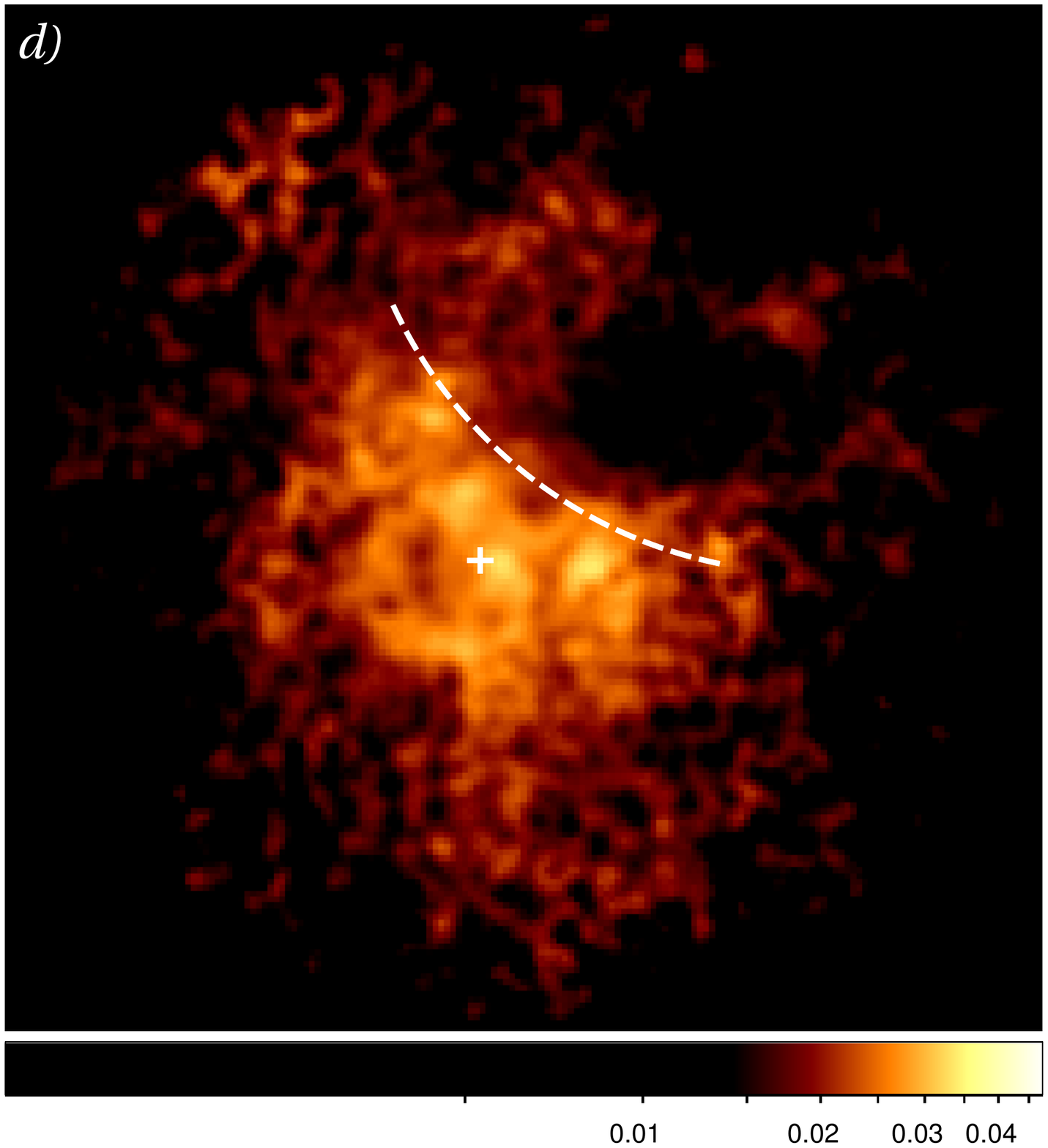}{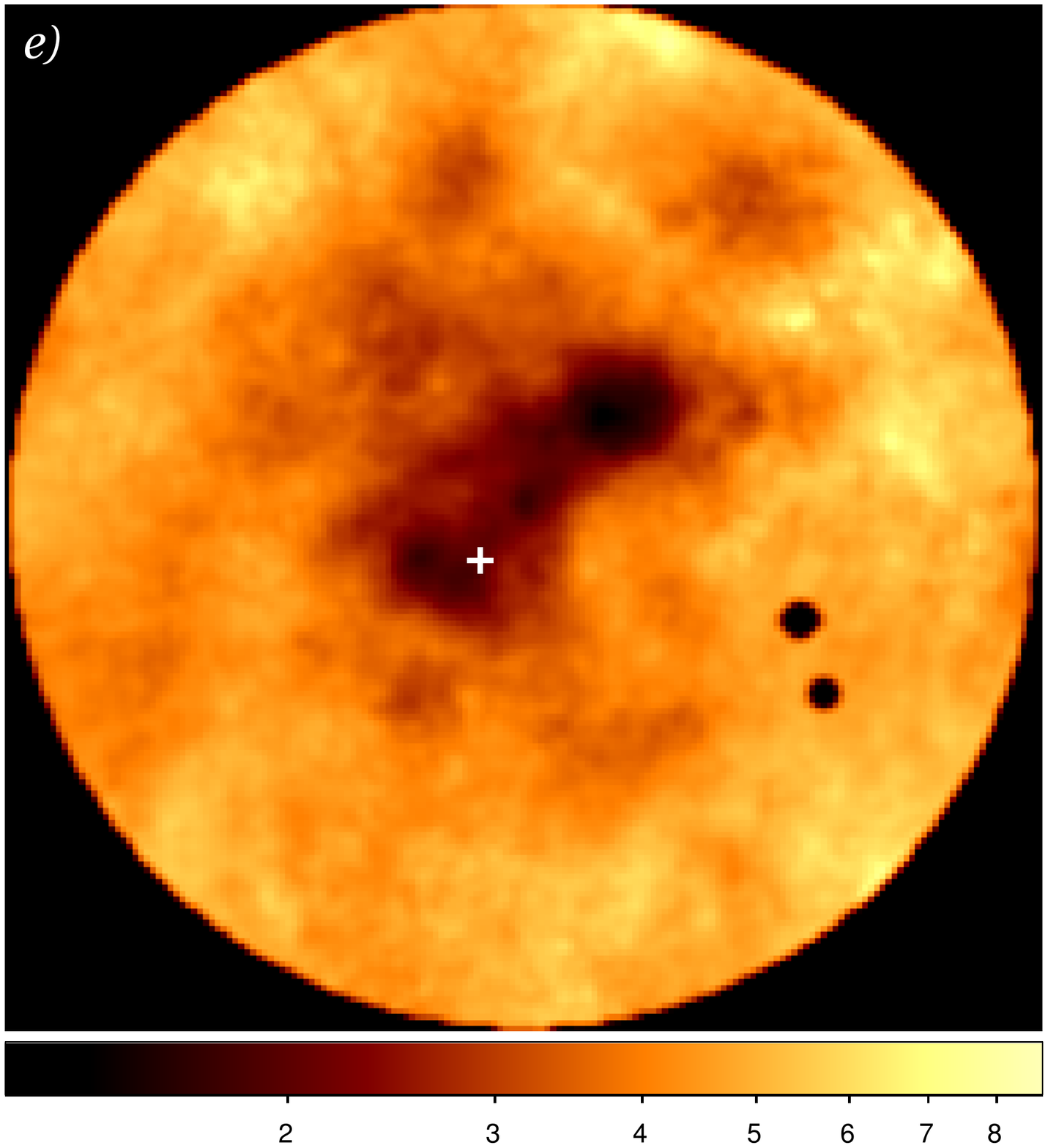}{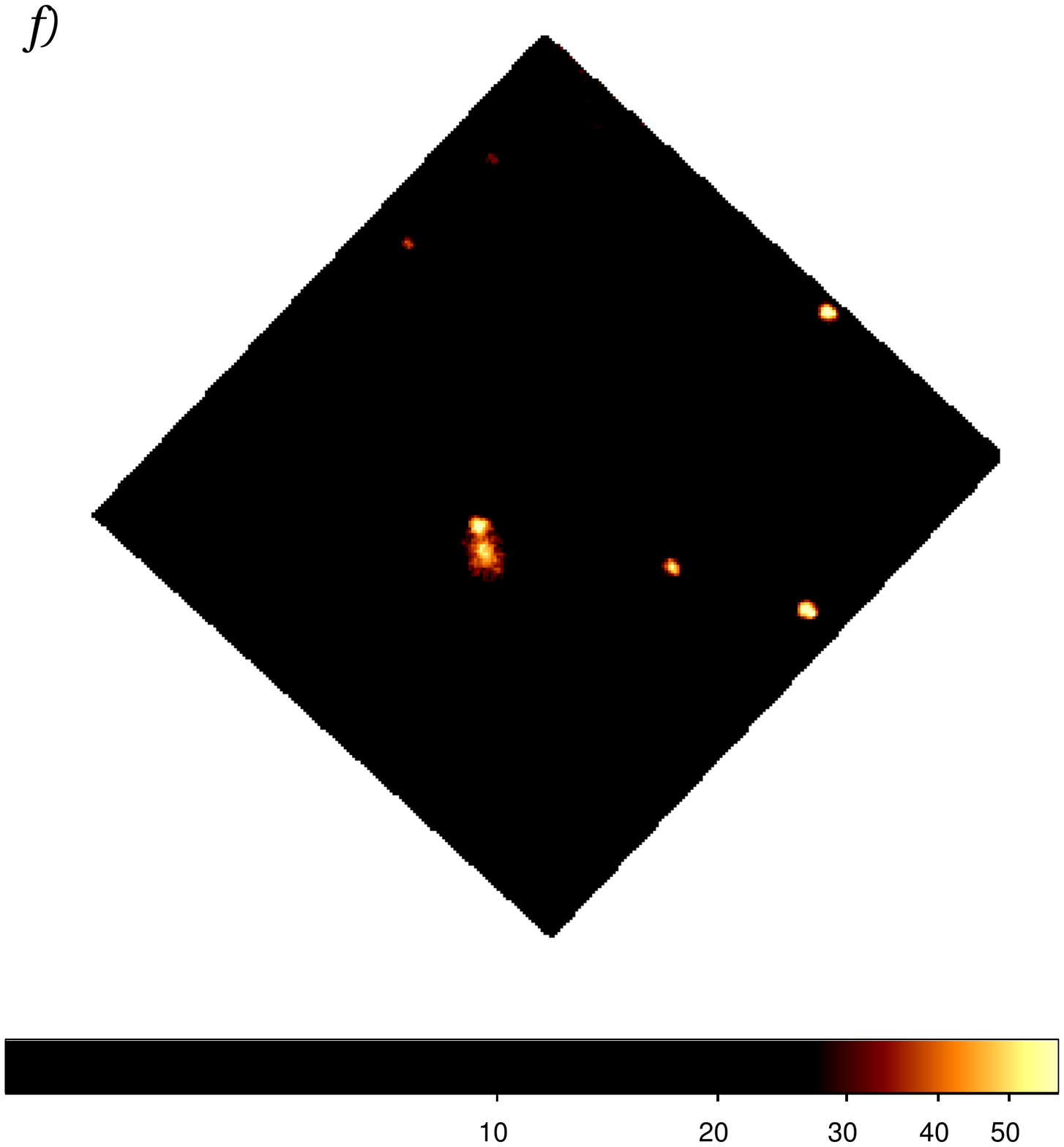}
\caption{
  Several projected maps of the core of Abell~133, matched in scale.
  The two dark spots seen to the southwest in panels {\it b} through
  {\it e} are
  source regions that were excluded in the X-ray analysis.  
  All images except panel {\it f} have been smoothed with a 2\arcsec\
  radius Gaussian.
  The white
  crosses mark the optical position of the cD galaxy.
  {\it a)} The 0.6--5.0~keV X-ray image, shown on the same scale for
  comparison.  The dashed line indicates the northeastern edge of the
  X-ray wings features.
  {\it b)} A higher resolution X-ray temperature map of the core of
  Abell~133.  The color-bar gives the temperature in keV.
  {\it c)} Abundance map, derived in the same manner as the
  temperature map, but with 4000 net counts per extraction region.
  The color-bar gives the abundance relative to solar.
  {\it d)} Pseudo-pressure map, in arbitrary units, derived from the
  projected density and 
  temperature maps.  The dashed line indicates the northern edge of
  the X-ray wing features.
  {\it e)} Pseudo-entropy map, in arbitrary units, derived from the
  projected density and 
  temperature maps.
  {\it f)} Raw {\it XMM-Newton} optical monitor image of Abell~133.
  The cD galaxy is resolved into two distinct peaks, with the cD position
  (marked with a white cross in the other panels) centered on the southern peak.
  \label{fig:tmap_core}
}
\end{figure}

A higher-resolution temperature map of the core region is shown in
Figure~\ref{fig:tmap_core}b.   
The extraction radii 
ranged from 3\arcsec\ (3.3~kpc) in the core to 15\arcsec\
(16.3~kpc) in the faint regions near the edge of the map.
  The white
cross marks the optical position of the cD galaxy.
The map reveals complex thermal structure in the core.  
The beginning of the eastern channel of cooler gas seen in the
larger FOV temperature map shown in Figure~\ref{fig:tmap} can be seen to
the east.
The plume is
visible as an extension of cool gas with minima on the northern and
southern tips, a cool central minimum, and some filamentary structure
along the northeastern edge, which roughly coincides with the
filamentary structure seen in the 1400~MHz radio observations
(Figures~\ref{fig:rad_imgs}a~\&~\ref{fig:rad_overlay}).
The coolest gas is at the northwest tip of the plume, although we show
in \S~\ref{sec:dxspec} that this is a projection effect, and that the
gas in the southern tip is as cool or cooler.  The southern minimum is
offset from the cD galaxy, and contains two distinct ``sub-minima''
separated by about 6\arcsec\ (6.5~kpc).  The raw {\it XMM-Newton
  Optical Monitor} ({\it XMM-OM}) image, shown in
Figure~\ref{fig:tmap_core}f, shows two 
similarly separated peaks, although at an offset position and position
angle.  The double peaks are also seen in {\it 2MASS}, {\it GALEX},
and less clearly in {\it DSS} images.  The positions for the peaks
match in {\it XMM-OM}, {\it 2MASS}, {\it GALEX}, and {\it DSS} images.
The positions of {\it Chandra} point sources in the field
match those from observations at other wavelengths, suggesting that the
astrometry is sound.  The temperature map is effectively smoothed due
to the extended size of the extraction regions, but the regions are
small near the core ($\sim$3\arcsec\ radii), and if the offset between
the optical and cool gas peaks was due to smoothing inherent in the
temperature map we would expect the gas peaks to be offset along the
cool plume of emission, to the northwest, whereas the gas is offset
southeast of the cD.  We conclude that this offset is real, and is
likely the result of a dynamical disruption of the core.  We discuss
this further in \S~\ref{sec:dynamics}.

The projected abundance map is shown in Figure~\ref{fig:tmap_core}c.
It was derived using the same method as the temperature map, but with
4000 net counts per extraction region, so that the abundances could
be measured to within 5-10\%.  The regions with highest abundance are
on either side of the plume, which has roughly half the abundance of
the surrounding region.  In \S~\ref{sec:dxspec}, we show that the
apparent lower abundance in the plume is a projection effect,
resulting from fitting a single temperature model emission from gas at
a range of temperatures.  Both the temperature and abundance maps show
regions of rapid change, or ragged edges, surrounding the central
plume at radii ranging from about 20 to 40~kpc.  These features are
examined in detail in \S~\ref{sec:profs}.

The projected pseudo-pressure and pseudo-entropy maps are shown in
Figures~\ref{fig:tmap_core}d~\&~\ref{fig:tmap_core}e.  
The column density map $n_{\rm col}$ was estimated using the square
root of the 1.2-2.5~keV {\it Chandra} image, as in Forman \etal\ (2007).
Combining this with the temperature map $kT$
we compute the pseudo-pressure map as $P_s = n_{\rm
  col} \, kT$ and the pseudo-entropy map as $S_s = kT / n_{\rm col}^{2/3}$.
The structure in the pseudo-entropy map is similar to what we see in
the temperature map.  The pseudo-pressure map shows an edge northwest
of the cD galaxy, which coincides with the northern edge of the wing
features in the X-ray image (compare the overlaid curve in
Figures~\ref{fig:tmap_core}a~\&~\ref{fig:tmap_core}d).  
We discuss this edge further in \S~\ref{sec:wings}.

To look for large-scale temperature structure, we generated a
coarsely binned (20\arcsec~pix$^{-1}$) temperature map of the entire
ACIS-I FOV from the most recent {\it Chandra} observation.  The result
is shown in Figure~\ref{fig:acisi_tmap}.  The extended emission
appears to be hotter on average in the south and southeast as compared to the
north, consistent with Figure~\ref{fig:tmap}.  There is a hot,
$\sim$6~keV spot to the southwest, and a cool clump of gas
extending to the edge of the chip to the north.  These features are
explored further in \S~\ref{sec:dxspec}.

\subsection{Detailed Spectra}  \label{sec:dxspec}

Based on the derived temperature map and X-ray image of the core, we
defined 6 regions for detailed spectral analysis of the core, shown
overlaid on the X-ray image and temperature map in
Figure~\ref{fig:regs}.  A summary of the fitted spectral models for
each region is given in Table~\ref{tab:xspec}, along with a fit to the
total diffuse emission within 1.9\arcmin\ (121~kpc) of our adopted
central position.  Each region was fitted with either a two
temperature {\sc apec} model (with abundances tied) or an {\sc apec}
plus power-law model.  In each case, the two component model was a
significant improvement over a single thermal model.  The fit to the
total diffuse emission implies an absorption of $N_H =
2.9^{+0.3}_{-0.3}\times 10^{20}$~cm$^{-2}$, above the Galactic value
of $N_H = 1.57\times 10^{20}$~cm$^{-2}$ derived from HI maps (Kalberla
\etal\ 2005).  This is consistent with what was found by F02.  Such
``absorption excesses'' have been found from X-ray observations of
other clusters (\eg, Pointecouteau \etal\ 2004).  Since the absorption
excess is likely due to variations in the Galactic absorbing column,
which can vary across the field in a complicated way, we left the
absorption free to vary in the fits in Table~\ref{tab:xspec}.  We
checked the impact of this freedom by redoing each fit with the
absorption fixed at the value found from the fit to the total diffuse
emission and at the Galactic value.  In all cases the best fitting
parameter values agreed well within the 1$\sigma$ confidence ranges,
except for the abundance of the low temperature component for the fit
to the total diffuse emission (which was 40\% larger).  We conclude
that either fixing the absorbing column at a reasonable value or
allowing it to vary does not significantly affect the fitting results
in Table~\ref{tab:xspec}.

The fits to regions R1-R3 find a higher abundance than that indicated
by the projected abundance map shown in Figure~\ref{fig:tmap_core}c.
This is due to the two temperature thermal model, which effectively
deprojects emission from the central core and plume.  Fitting a
multi-temperature spectrum with a single thermal model, as is done
when generating the temperature map of the core, will generally give
an abundance measurement that is lower than the true
abundance of the gas, although the best-fitting temperature is expected
to be within the range bounded by the lowest and highest temperature
components (see Buote \etal\ 2000).  Similarly, the detailed spectral
fits show that the 
temperature of the cool
gas in the southern peak (R1), near the cD galaxy, is consistent with the
temperature of the gas in the ``cap'' of the plume (R3).  This is in contrast
to the temperature map shown in Figure~\ref{fig:tmap_core}b, which
shows the coolest gas in the cap.  This suggests that projection
effects from hotter gas at larger radii is more significant for the
southern peak than the northern cap.

\subsubsection{X-ray Spectrum of the Radio Relic}\label{sec:xrelic}

The spectrum of the region containing the radio relic in the core of
Abell~133 is of particular interest as we expect some non-thermal
IC
emission from the interaction between CMB
photons and the synchrotron-emitting particles in the relic.
Additionally, the relic may contain thermal gas that is so hot and
tenuous that its X-ray emission is negligible, although it contributes
significantly to the pressure support for the X-ray cavity
(B\^{i}rzan \etal\ 2008 find that, under the assumptions of
equipartition and cavities devoid of X-ray emitting gas, the ratio of
energy in protons to electrons $k$ must be several hundred to
achieve pressure balance with the surrounding ICM).
Region R4 roughly corresponds to the region of the radio relic shown
in Figure~\ref{fig:rad_imgs}, and to the relic region defined by F04.
As shown in Table~\ref{tab:xspec}, the two temperature and temperature
plus power-law models both provide an improved fit over the single
temperature model.
For the two temperature model, the higher temperature could not be
significantly constrained.
However, if we assume that the hot gas is contained within the radio
relic, we can place an upper limit on its temperature by requiring
that the thermal pressure be less than the local thermal pressure
outside of the relic.  Assuming an oblate spheroidal geometry for the
relic with the minor axis in the plane of the sky, and taking the
external pressure from the deprojected profiles presented in
\S~\ref{sec:profs}, the lower limit on the normalization of the high
temperature component suggests an upper limit of $kT \la 6$~keV.
For the thermal plus power-law model, using the best-fit photon
index, we find a 0.3-10~keV flux from the
power-law component  
of $F_{\rm PL} = 1.3 \times 10^{-13}$~ergs~cm$^{-2}$~s$^{-1}$
with an upper limit of $F_{\rm PL} < 3.1 \times
10^{-13}$~ergs~cm$^{-2}$~s$^{-1}$, which is lower than the previous 
tightest constraint of $F_{\rm PL} < 5.4 \times
10^{-13}$~ergs~cm$^{-2}$~s$^{-1}$ derived by F04 from {\it XMM-Newton}
observations.  
We note that the best-fitting photon index is relatively
flat, and inconsistent with the steep indices found in this region
from the radio observations (see ~\S~\ref{sec:sindex}).  If this
component represented a detection of IC emission from the radio
emitting particles we would expect the indices to be roughly equal.
However, although the 1$\sigma$ confidence intervals do not overlap
the radio and X-ray indices agree within the 90\% confidence intervals
(although with large uncertainties on the X-ray index).
We conclude that statistically we cannot distinguish
between a high temperature thermal model and a non-thermal model
for the second component in the region of the relic.  Although the
flat spectral
index argues against this component arising solely from IC emission from the
radio-emitting particles, the errors are large, and the X-ray and
radio slopes are plausibly consistent.

\subsubsection{The Double Core}\label{sec:dcore}

As noted in \S~\ref{sec:tmap}, Figure~\ref{fig:tmap_core}b reveals two
distinct temperature minima within the core.
The double-peaked nature of the core is
also evident in several other observations at other wavelengths,
although the cool
gas peaks are slightly offset to the southeast.
The multiple peaks may represent an early stage galaxy-galaxy merger
within the cD galaxy.
To determine whether the gas in each peak has a common
origin, we fit spectra extracted from each region (R5 and R6 in
Table~\ref{tab:xspec}). The temperatures are the same, and although
the best-fitting abundance value is substantially higher in the
northeastern minimum, they are in agreement within the (large) errors.
Assuming a spherical geometry for each region and calculating the entropy as
\begin{equation}
S = kT/n_e^{2/3},
\end{equation}
we find that in region R5 $S_{\rm R5} = 10.4^{+3.2}_{-8.1}$~keV~cm$^2$
and for R6 $S_{\rm R6} = 5.28^{+1.0}_{-0.5}$~keV~cm$^2$ (errors are
statistical only, and do not include uncertainties due to assumptions
about geometry, etc.).  Although the entropy of the gas in the
optically brighter, southwestern
minimum is substantially lower, the values are consistent within the errors.
We conclude that, although the abundance and entropy measurements are
suggestive, we cannot 
unambiguously identify these structures as originating from distinct systems.

\subsubsection{Thermal Structure of the Wings and Plume}\label{sec:xplume}

F04 find evidence for a temperature gradient along the plume, such
that the temperature decreases from the core in the southeast to the
cap in the northwest.  
They interpret this as support for their model where the wing features
shown in Figure~\ref{fig:xr_core} are the result of a merger shock
propagating through the cool core from the southeast to the northwest.
The front is distorted as it passes through the denser gas
in the core, giving it a curved appearance.
Our spectral fits to R1, R2, and R3 show no evidence for
temperature differences.  To further test for changes in temperature
along the plume,
we split the plume into 7 contiguous regions, 
including the core and northern cap, each containing about 2000 net
counts.  Each region was fitted with a two temperature model, with
abundances tied and fixed at 1.3 times solar and the higher
temperature fixed at 2.3~keV, consistent with results from the fits to
R1-R3 in Table~\ref{tab:xspec}.  The absorbing column density was
fixed at $N_H = 2.9 \times 10^{20}$~cm$^{-2}$.  
We find no evidence for systematic temperature variations along the plume.
All regions are consistent with a lower temperature of 1.1~keV
within 3$\sigma$, and most are consistent with this temperature to
within 1$\sigma$.

F04 also find marginal evidence for a temperature jump across the edge
defined by the wing features (indicated by the white dashed line in
Figure~\ref{fig:tmap_core}), consistent with a shock passing through
the core from the southeast.  This result relies on fitting a two
temperature model and fixing the abundances and temperature of the
hotter component.  Their single temperature models show no temperature
jump across the edge (see their Table~4).  To test this result with
the new {\it Chandra} data, we defined regions corresponding to those
shown in F04 (their Figure~12).  The results from fits to these regions
are given in Table~\ref{tab:wings}, where we have adopted the labels
from F04 of WU, WD, NU, and ND (west upstream, west downstream, etc.,
where ``upstream'' and ``downstream'' refer to the position relative
to the putative shock front).  For the {\it Chandra} data we consider
here, we have sufficient counts in each region to fit two temperature
models in each region, and therefore directly measure projection
effects.  We do not reproduce the temperature rise in the upstream
region reported by F04, and in fact find the upstream region to be
cooler in the northeastern wing.  We discuss this discrepancy further
in \S~\ref{sec:wings}.

\subsubsection{Thermal Structure of the Extended ICM}\label{sec:xlfov}

As with the core region, we followed up features seen in the large FOV
temperature map shown in Figure~\ref{fig:acisi_tmap} with detailed
spectral analysis.  
The map shows a cool clump of gas 8.7\arcmin\
(564~kpc) to the northeast, a 
hot spot 7.2\arcmin\ (468~kpc) to the southwest, and a large hot trail of
emission south of
the core.  We defined four regions (labeled in
Figure~\ref{fig:acisi_tmap}), one each for the hot spot and northern
clump, and, since the northeast and southwest are contaminated by the
clump and the hot spot, one each for the southeast and northwest
corners to test for temperature asymmetries.  The results are shown in
Table~\ref{tab:acisi_spec}.  
Since the features in question are in relatively faint regions, we
checked the impact of systematic uncertainties in the background by 
varying the background normalization by $\pm5$\% and re-fitting the
spectra.  In no case were the results significantly affected, with the best
fitting values agreeing well within the 1$\sigma$ confidence ranges.

The northwestern and southeastern corners
are well described by single temperature models with temperatures of
4.1~keV and 4.6~keV respectively (adding a second
thermal or power-law component did not improve the fit in either
case).  The detailed fits support the impression from the temperature
map that the emission is hotter to the south, with the temperature
difference significant at 2.5$\sigma$.  The northern clump is also
well-described by a single thermal model at a much lower temperature
of 2.3~keV.  Adding a second thermal component did not improve the
fit, suggesting that the contribution from the projected hot ICM is
negligible.  

The hot spot to the southwest is reasonably
well-described by a single temperature model with $kT = 6.7$~keV.  If
a second thermal or power-law component is included the fit is
slightly improved, as measured by the chi-squared per
degree-of-freedom, $\chi^2_\nu$, with the null hypothesis probability
increasing from 30\% to 45\%.  In both cases the lower temperature
component is roughly 2.5~keV.
In the case of the two temperature
model, the higher temperature is only very poorly constrained, with
$kT_{\rm hot} \ga 11$~keV.  Both models describe the data equally
well, and we cannot discriminate between them.

As a check on these results, we extracted spectra from each of the
regions shown in Figure~\ref{fig:acisi_tmap} from the {\it XMM-Newton}
observations presented in F04.  We find temperatures of $kT_{\rm SE} =
3.03 \pm 0.09$, $kT_{\rm NW} = 2.83 \pm 0.1$, $kT_{\rm N. Clump} =
1.89^{+0.57}_{-0.24}$, and $kT_{\rm Hot Spot} =
2.72^{+0.49}_{-0.34}$.  Thus, we confirm results from the {\it
  Chandra} measurements (Table~\ref{tab:acisi_spec}), with the
emission in the southeast being somewhat hotter than the emission in
the northwest by about 10\% (at 1.5$\sigma$ significance), and the
northern clump being cooler than both (at 1.6$\sigma$ significance).
However, we do not confirm the higher temperature of the hot spot found
from {\it Chandra} observations.  Additionally, the measured {\it
  XMM-Newton} temperatures are significantly lower than the {\it
  Chandra} temperatures for the same regions
(Table~\ref{tab:acisi_spec}).  This latter result may be due to
cross-calibration issues (although our temperature measurements from
{\it Chandra} and {\it XMM-Newton} agree well in the region of the
bright core), 
or it may be due to contamination from the
bright cool core (which has temperatures in the 2--3~keV range) as a
result of
{\it XMM-Newton's} relatively large PSF.  The fact that the hot spot is
not detected in the {\it XMM-Newton} data may also be a result of the
large PSF, with scattered emission from brighter, cooler regions swamping
the emission from the hotter (possibly non-thermal) component.
However, it is possible that the high {\it Chandra} temperature of the
hot spot is a spurious result, possibly due to systematic uncertainty
related to the background or instrumental calibration.
Thus, we conclude that, while the measured temperature differences are
only marginally significant, the results from {\it XMM-Newton} support the
results from {\it Chandra} on the relative temperatures of the
northwestern and southeastern corners and the northern cool clump.
We do not find evidence for the southwestern hot spot in the {\it
  XMM-Newton} data.

\subsection{Deprojected Profiles} \label{sec:profs}

In this section, we consider the average radial temperature, abundance, and
density profiles of the gas in Abell~133.  Accurate deprojected profiles out to
large radii have already been given by Vikhlinin \etal\ (2005) using
{\it ROSAT} and the earlier {\it Chandra} data.  We
therefore only consider the combined {\it Chandra} data here, with the
focus on detecting structure in the central regions, which may gives
clues as to the dynamical state of Abell~133.

Figure~\ref{fig:edges_img} shows a smoothed {\it Chandra} image of the
core, with smoothing and scaling chosen to highlight structure in the
diffuse emission.  The innermost cool gas, which makes up the core and
the plume, is sharply contrasted with the surrounding gas,
particularly to the southeast, where there is a prominent edge feature.
Additionally, there is an outer elliptical edge which completely
encompasses the central core.  This corresponds to the edge detected
in the southeast by F02, with the deeper {\it Chandra} observation
revealing its full annular extent.  To the northwest, the edge lies
just beyond the tip of the plume.  Our adopted center for radial
profiles, indicated by the blue cross in Figure~\ref{fig:edges_img},
is roughly at the center of the outer elliptical edge.  The outer edge
roughly corresponds with the edges in the temperature and abundance
maps (Figures~\ref{fig:tmap_core}b,c) noted above.

To quantitatively study these edges, we extracted the
0.6-5.0~keV surface brightness profile in
elliptical annuli, centered on the blue cross 
in Figure~\ref{fig:edges_img}.  Each annulus had an axial ratio of
$a/b=1.39$, where $a$ and $b$ are the semi-major and -minor axes,
respectively, and a major axis position angle of 20\mydeg\ from north,
which roughly corresponds to the shape of the edge feature indicated
in Figure~\ref{fig:edges_img}.  The sector covering the plume feature, 
between 285-352\mydeg\ measured east from north, was excluded.  

The resulting integrated emission measure profile is shown in
Figure~\ref{fig:emprof}. 
The profile has a power law shape, with
discontinuous jumps at roughly 20 and 30~kpc (where distance is
measured along the minor axis of the ellipse, such that the jumps
are at 28 and 42~kpc along the major axis).  Following our previous
work, we fit the profile by
projecting a 3 dimensional density profile consisting of three power
laws, connected with two discontinuous breaks, or ``jumps''.  The free
parameters were the normalization, the inner ($\alpha$), middle
($\beta$), and outer ($\gamma$) power law slopes, the position of the
density discontinuities ($b_{\rm 
  break,1}$ and $b_{\rm break,2}$), and the amplitude of the jumps
($A_1$ and $A_2$).  A prolate spheroidal geometry was assumed, with
the major axis in the plane of the sky.  The
best-fitting model is shown as the solid line in
Figure~\ref{fig:emprof}, with the fitted break radii indicated by the
vertical dashed lines.  The model provides an acceptable fit, with
$\chi^2_\nu = 16.4/11 = 1.5$. 
This is a significant improvement over the fit from a single power law
model, which gave $\chi^2_\nu = 94.6/17 = 5.6$.  The best-fitting
parameters (with 90\% confidence intervals) were $\alpha =
-0.40^{+0.05}_{-0.10}$, $\beta = -0.72^{+0.12}_{-0.16}$, $\gamma =
-1.19^{+0.02}_{-0.02}$, $b_{\rm break,1} = 18.7^{+0.37}_{-0.75}$~kpc,
$b_{\rm break,2} = 31.9^{+1.28}_{-1.28}$, $A_1 =
1.29^{+0.05}_{-0.03}$, and $A_2 = 1.18^{+0.02}_{-0.02}$.  The fitted
model therefore confirms the presence of small, but statistically
significant, density jumps at roughly 19 and 32~kpc along the minor
axis of the elliptical extraction region, coincident with the surface
brightness edges indicated in Figure~\ref{fig:edges_img}.

We also extracted deprojected density, temperature, and abundance
profiles using the ``onion peeling'' method employed, \eg, by Blanton
\etal\ 2003.  First, the projected temperature,
abundance, and {\sc xspec} normalization are determined by fitting an
absorbed {\sc apec} model to the outermost annulus.  
Fits to spectra from annuli at smaller radii are then determined by
adding an additional component for each outer annulus, with fixed
temperature and abundance, and a normalization scaled to project from
the outer to the inner annulus.  
We note that this procedure misrepresents the errors, since
the contributions from outer shells are fixed.
The absorbing column density was fixed at $N_H =
2.9 \times 10^{20}$~cm$^{-2}$.
As above, a prolate spheroidal
geometry was assumed, using annuli with the
same axial ratio and position angle as those used to extract the
emission measure profile shown in Figure~\ref{fig:emprof} (again, the
sector containing the plume was excluded).  

The resulting deprojected
electron density profile is shown in Figure~\ref{fig:neprof}.  The
dashed lines indicate the best fitting break radii derived above from
the emission measure profile.  The profile is consistent with small
jumps in the vicinity of the break radii, although the inner jump is
offset by a few kpc (\ie, a few arc-seconds).  The deprojected
temperature profile (dotted lines) is plotted along side the projected
temperature profile (solid lines) in Figure~\ref{fig:ktprof}.  Both
temperature profiles show a rapid increase across the inner edge and a
leveling off just outside the edge, from 2.69 to 3.07~keV in the
deprojected profile.  The deprojected temperature profile is
consistent with a similar temperature jump across the outer edge, from
3.07 to 3.33~keV.
Finally, the projected and deprojected abundance profile is shown in
Figure~\ref{fig:abprof}.  The deprojected profile indicates a
decrease in abundance across the inner edge, with no obvious
jump across the outer edge.  The inner abundance edge is qualitatively
similar to edges seen in M87 reported by Simionescu \etal\ 2007, who
conclude that they may indicate bulk, or, in the case of M87,
oscillating motion of the central cD galaxy with respect to the
ambient ICM.
From the temperature and abundance jump
across the inner edge, we conclude that this feature represents the
interface between the cool, metal-rich gas associated with the central
cD galaxy and the hotter surrounding gas.
The outer edge has characteristics of a cold front, now commonly seen
in galaxy clusters, where a temperature increase across a surface
brightness edge is accompanied by a corresponding density decrease,
such that the pressure profile is continuous across the edge (see Markevitch \&
Vikhlinin 2007 for a review), although, as seen in
Figure~\ref{fig:tmap_core}b, the structure of the edge is ragged and
somewhat complicated, unlike the smooth edges seen in other systems.
This feature may represent a sloshing cold front, as seen in
simulations (Ascasibar \& Markevitch 2006) and observations of other
systems (Mazzotta \etal\ 2001; Dupke \etal\ 2007; Gastaldello \etal\
2009; Randall \etal\ 2009), possibly disrupted by the activity
associated with the radio relic (resulting in the formation of the
plume and the X-ray cavities) and viewed at an angle such that the
oscillatory motion is not close to the plane of the sky.  We discuss
this feature in the context of the dynamical state of Abell~133
further in \S~\ref{sec:dynamics}.

\section{Radio Spectra} \label{sec:rspec}

\subsection{Spectral Index Maps} \label{sec:smaps}

Using the radio observations at multiple frequencies, we constructed
2D spectral index maps of the features seen in Figure~\ref{fig:rad_imgs}.
The radio spectral index $\alpha_{\rm r}$ relates the integrated
brightness $S$ and frequency $\nu$ by
\begin{equation}\label{eq:rad}
S(\nu) \propto \nu^{\alpha_{\rm r}}.
\end{equation}
To study the low frequency spectral index structure we
have cut the UV coverage to be matching for the A configuration 74~MHz
and B configuration 330~MHz data sets. After re-imaging with the new
UV ranges, we convolved the (similar resolution) 74 and 330~MHz images
to a circular beam of $\theta$ = 42\arcsec. Finally, the two input
maps were blanked at the 5$\sigma$ level on each map prior to
constructing the spectral index map of the target. In addition to the
spectral index map, we have calculated an error map based on the
statistical noise in the images. We account for
calibration errors in calculating the final error on the spectral
index by including an uncertainty of 6\% on the flux calibration of
the 74 MHz data (Cohen \etal\ 2004) and a 3\%
calibration error on the 330 MHz data (Lane \etal\ 2004). We use
similar techniques to those described above to obtain spectral index
maps between 330 and 1400~MHz for both the high and lower resolution
data sets.

The resulting spectral index maps are shown in Figure~\ref{fig:index_maps}.
The structure of the maps support the impressions given by the radio
brightness maps presented in \S~\ref{sec:rimg}: the structure to the
north consists of a relatively flat ($\alpha_{\rm r} \ga -1$) core adjacent to a
steep ($\alpha_{\rm r} \approx -2$) diffuse source, which has
previously been classified as a radio relic (S01).  The
southern core is also found to be flat, with two steeper jets
extending to the north and south.  In Figure~\ref{fig:index_maps}c we
see that the global structure is consistent with what would be
expected from a double-lobed radio source: a flat core (the southern
core labeled in Figure~\ref{fig:index_maps}a) with two jets that
steepen with distance and terminate in lobes.  The northern lobe is
coincident with the radio relic in the core of Abell~133, while the
southern lobe is steeper in the west where it connects to the jet (we
verified that this gradient is not due to contamination from the
nearby {\it NVSS} source).
The spectral maps are therefore consistent with the
interpretation suggested in \S~\ref{sec:rimg}, where emission from a
distant, background radio galaxy associated with the southern core
overlaps emission from the radio relic in Abell~133.

\subsection{Integrated Spectral Index of Individual Features} \label{sec:sindex}

In addition to the spectral maps, we calculated the average integrated
spectral index and measured fluxes for some of the main features.  The
results are given in Table~\ref{tab:sindex}.  
Our flux and spectral index measurements for the relic/northern lobe
in Abell~133 are consistent with results given by B\^{i}rzan \etal\ (2008).
Both the southern core and the core source in the Abell~133 cD galaxy
have relatively steep spectra for cores, with indices $\ga -1.2$.  This
suggests either that each source is a ``dying'' core, or, more likely, that
contamination from nearby diffuse emission is steepening the
indices.  The spectral index maps seem to support contamination:
for example in the 330 -- 1400~MHz map in the region of the
southern core, the indices get as flat as 
-0.8, and for the Abell~133 core they reach -0.4.  These peak
values are slightly offset from the cores in a direction opposite from
the brightest diffuse emission, suggesting that the diffuse emission
contaminates the spectral index measurements at the location of the
cores but has a smaller effect on the side of the core where the diffuse
emission is fainter.  The measurements of the southern core, its jets,
and the arc show the spectral index steepening with distance, although
these regions may suffer from missing flux at 1.4~GHz and are
therefore artificially steep.

\section{Discussion} \label{sec:discuss}

\subsection{The Dynamical State of Abell~133} \label{sec:dynamics}

It is clear that Abell~133 has been dynamically
disturbed.  At smaller radii, the surface
brightness map shows an elliptical edge surrounding the core between
30-50~kpc (Figure~\ref{fig:edges_img}), which is roughly aligned with
the ellipticity at larger radii.  The temperature map
(Figure~\ref{fig:tmap_core}b), temperature profile
(Figure~\ref{fig:ktprof}), and density profile (Figure~\ref{fig:neprof})
show evidence for modest jumps at the location of the edge.  The edge
is well-described by a spheroidal discontinuous power-law
density model, with a density jump of about 1.18 (\S~\ref{sec:profs}),
consistent with what is expected from gas sloshing in the core due to
interactions with a merging subgroup.  The temperature map of
the core (Figure~\ref{fig:tmap_core}b) shows two distinct temperature
minima within the central cD galaxy, indicative of some small-scale
dynamical process (possibly a galaxy-galaxy merger).  Similar double
peaks are seen in optical, UV, and NIR images, although the position and
rotation angle of the X-ray minima are offset from these features.
This offset also suggests a dynamical process acting on the gas, such
as gas sloshing or stripping, where the cool
diffuse gas clouds have been displaced from the potential minima of
their respective galaxies, while the collisionless stars remain
undisturbed.  Thus, although the evidence for gas sloshing from the
X-ray edges alone is not as conclusive as seen in some other systems
(\eg, NGC~5098, Randall \etal\ 2009), we conclude that there is strong
evidence that Abell~133 has been dynamically disturbed by an
interloping subgroup, and we suggest that the elliptical edge in
the surface brightness map is a gas sloshing edge, seen in simulations
and in other galaxy groups and clusters, possibly viewed at a high
angle of inclination such that the plane of the sloshing motion is not
close to the plane of the sky, leading to the somewhat ragged
appearance of the edge.

Since gas sloshing requires a relatively massive perturber, we looked
for evidence of substructure in Abell~133.
The {\it RASS} contours shown in Figure~\ref{fig:fullimg} indicate a
peak roughly 10.6\arcmin\ (690~kpc) southwest of Abell~133.  This peak
is outside the FOV of the {\it Chandra} images.  The {\it XMM-Newton}
data reveal a bright source at this location, with a spectrum that is
reasonably well-described 
by a power-law (with photon index $\Gamma = 2.33 \pm 0.03$ and
$\chi^2_\nu = 154/108 = 1.4$), and 
poorly described by an {\sc apec} model (with $\chi^2_\nu = 344/107 = 3.2$).
This source is therefore likely an AGN.  Thus, we see no evidence for
a nearby subcluster in the available X-ray data.

Using optical data and redshift measurements, Way \etal\ (1997) find
evidence for substructure in Abell~133, including an associated
subgroup to the southwest, and conclude that Abell~133 is likely
``dynamically young''.  Using a 2D wavelet analysis technique, Flin \&
Krywult (2006) find evidence for three distinct substructures within
1.5~Mpc of the cluster center, all roughly along the major axis of the
elliptical X-ray brightness contours of Abell~133,
at 1970~kpc and 550~kpc from the X-ray core (we note that their
adopted cluster center for Abell~133 corresponds roughly to the southwestern
source seen in the {\it ROSAT} image, which is why it is possible for
the separation from the X-ray center to be larger than their maximum 1.5~Mpc
search radius).  Finally, Dalton \etal\ (1997) report an optically
selected galaxy cluster at the position of the {\it NVSS} source
indicated in Figure~\ref{fig:rad_imgs}d at $z = 0.06$, which is close to the
redshift of Abell~133 ($z = 0.0566$) and likely an associated subcluster.

As a further check for evidence of substructure, we performed our own
3D substructure test, using published galaxy positions and redshifts
and a technique based on Kolmogorov-Smirnov (K-S) likelihood analysis that
we have employed previously (\eg, Owers \etal\ 2009).  For our sample,
we used all galaxies within 3.5~Mpc with redshifts given in the
NASA/IPAC Extragalactic Database (NED), plus those given in Way \etal\
(1997).  The ``bubble'' plot for the resulting 138 galaxies is shown
in Figure~\ref{fig:bubble}.  In this plot, each circle has a radius 
\begin{equation}\label{eq:bubble}
r \propto -\log{P},
\end{equation}
where $P$ is the probability (measured using the K-S likelihood) that the
peculiar velocity distribution
of the $\sqrt{N}$ nearest neighbors to the galaxy is the same as
the global velocity distribution (where $N = 138$ is the total number
of galaxies in the sample).  Blue/red circles have negative/positive
peculiar velocities relative to the cluster redshift, and bold
circles occurred less than 10\% of the time in 5000 Monte Carlo
simulations.  The total k-statistic, which is a measure of substructure
in the cluster, is given by the sum of $-\log{P}$ for the entire sample.
We find that the total k-statistic measured is not significant, and is
in fact only 0.03$\sigma$ from the mean of the distribution produced by 5000
Monte Carlo realizations, where the velocities are randomly shuffled
between galaxies.  
Overlaid are galaxy density contours selected from the {\it
  SuperCOSMOS} survey (Hambly \etal\ 2001).  Cluster members were
selected by requiring that they lie within $\pm0.2$~mag of the
red-sequence to a limiting magnitude of $r_F=20$.  The red-sequence
was defined by matching the spectroscopically 
confirmed cluster members with the {\it SuperCOSMOS} data and fitting the
red-sequence with a straight line. There were 792 galaxies that met
this criteria. 

Despite the insignificant total k-statistic found from the bubble plot
analysis, Figure~\ref{fig:bubble} shows a
concentration of blue circles to the southwest (4 of which occurred
less than 10\% of the time in the Monte Carlo simulations) that is
roughly coincident with the substructure detected by Flin \& Krywult
(2006) and a
peak in the overlaid galaxy density contours.
Furthermore, there appears to be a similar excess of red-shifted
galaxies $\sim300$~kpc south of Abell~133.
The velocity histograms for all 138 galaxies, as well as those for the
southwestern and southern excesses seen in the bubble plot, are shown
in Figure~\ref{fig:bubble}.  
To confirm the peculiar velocity of the
cD galaxy with respect to the cluster mean, we calculated the ${\rm
  Z}_{score}$ as in Way \etal (1997) with our (slightly) larger
sample.  We find ${\rm Z}_{score} = 0.256^{+0.170}_{-0.162}$, with a
value less than zero (which would indicate a cD velocity consistent
with the cluster mean) occurring only 26 times in 10,000 bootstrap
re-samplings.  This is essentially the same result found by Way \etal\
(1997), confirming that the cD has a high peculiar velocity, as expected
in a sloshing scenario (Ascasibar \& Markevitch 2006).
Thus, although the ``bubble'' plot test is inconclusive,
results from previous studies 
argue for the presence of substructure in the cluster, with potential
subgroups $\sim 300$~kpc and $\sim600$~kpc south/southwest of the cluster
center, consistent with the major axis of the elliptical X-ray
isophotes, thus
providing perturber candidates to explain the dynamical disturbance of
Abell~133.

We can estimate
the time since core passage for a perturbing subcluster using the
total mass profile for 
Abell~133 derived by Vikhlinin \etal\ (2006).  They find that the
density profile is well described by a NFW profile (Navarro et
al. 1996, 1997) $\rho(r) \propto (r/r_s)^{-1}(1 + r/r_s)^{-2}$ with
$r_s = 317$~kpc, and that the total mass within 500 times the critical
density at the cluster redshift is $M_{500} = 3.2 \times 10^{14} \msun$.
Therefore, for a test particle starting at rest from a turn-around
radius of 3.6~Mpc (with a small impact parameter), the time since 
core passage for a current separation of 600~kpc (300~kpc) is $2.3 \times
10^8$~yr ($1.0 \times 10^8$~yr).  This is a lower limit for the
observed subclusters since they need 
not necessarily move in the plane of the sky, and the true
separation is likely to be larger than the projected separation.
Given the lower limit
on the age of the radio relic from B\^{i}rzan \etal\ (2008) of
$t_{relic} > 5.8 \times 10^7$~yr, it is plausible that one of the 
subclusters passed by the core during the lifetime of the relic,
affecting its morphology.  We discuss the implications of this
scenario for the radio relic in \S~\ref{sec:relic}.

The large FOV temperature map (Figure~\ref{fig:acisi_tmap}) and the
detailed spectral fits presented in Table~\ref{tab:acisi_spec} also
show some evidence for merger activity.  The elongated cool clump of
gas to the north, which is significantly cooler than the ambient ICM
at similar radii, may be the remnant of material that has been
stripped from a merging subcluster as it passed from the northeast
to the southwest.  The orientation is consistent with the orientation
of the elliptical X-ray isophotes in the diffuse emission
(Figure~\ref{fig:fullimg}).  Furthermore, the hot trail of gas south
of the cluster is similar to what would be expected if a subcluster
passed through this region, shock heating the gas as it passed from
the east, then south of the core.  This is consistent with the
structure of the inner X-ray surface brightness edges shown in
Figure~\ref{fig:edges_img}, in that the sharpest edge is to the
southeast, as would be expected from gas sloshing if the perturber
passed southeast of the core (see Ascasibar \& Markevitch 2006).  We
note that the interpretation of the cool clump to the north as
stripped gas from a subcluster and the hot trail to the south as shock
heated ICM gas from the passage of the subcluster are not necessarily
inconsistent, since the orbit of the subcluster may be curved as
projected onto the plane of the sky (although one might expect the
elongation axis of the cool clump to point more to the south in such a
unified model).  Finally, the hot spot to the southwest is
equally well-described by a two temperature model with
a poorly constrained high-temperature component at $kT \ga 11$~keV, or a
thermal plus power-law model.  Either of these are consistent with a
shock in this region, where the second component either detects shock
heated gas or inverse Compton (IC) emission from shock-accelerated particles.
However, the presence of this hot spot is not confirmed by {\it
  XMM-Newton}, and while this may be due to contamination by scattered
light from bright, cooler regions due to {\it XMM-Newton's} large PSF
we consider the {\it Chandra} detection of the hot spot to be a
marginal result.

From the combined evidence above, we propose a model where a merging
subcluster has passed from the northeast to the southwest of the core,
possibly leaving behind stripped gas to the northeast and shock
heating the ICM south of the core.  
The merger has elongated the X-ray isophotes at large radii in the
northeast-southwest direction.
The subcluster passed in front of
or behind the core, initiating sloshing in the central gas in a plane
at some inclination angle with respect to the plane of the sky.  Since the
putative spiral of sloshed gas is viewed from the side, it has an
elliptical appearance with ragged edges.
Optical observations indicate a subcluster roughly
600~kpc southwest of the core, which seems a plausible candidate for
the merging subcluster (a second optically selected subcluster 300~kpc
south of Abell~133 is also a viable perturber candidate, although it is
located too far to the east to explain the western hot spot shown in
Figure~\ref{fig:acisi_tmap} as shock heated gas).

\subsection{The X-ray Plume} \label{sec:plume}

\begin{figure}
\plotone{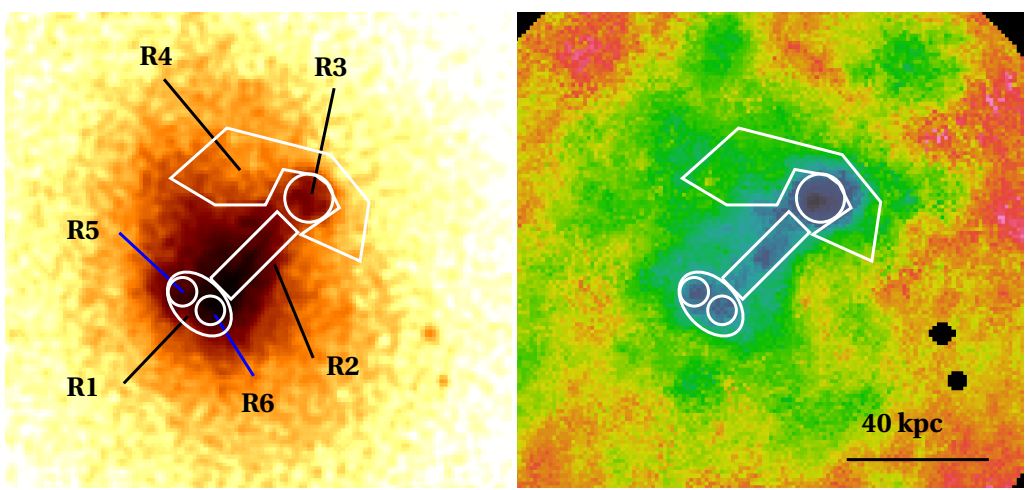}
\caption{
  Regions fitted in Table~\ref{tab:xspec} overlaid on the
  smoothed X-ray image ({\it left}) and unsmoothed temperature map
  ({\it right}).
\label{fig:regs}
}
\end{figure}

\begin{figure}
\plotone{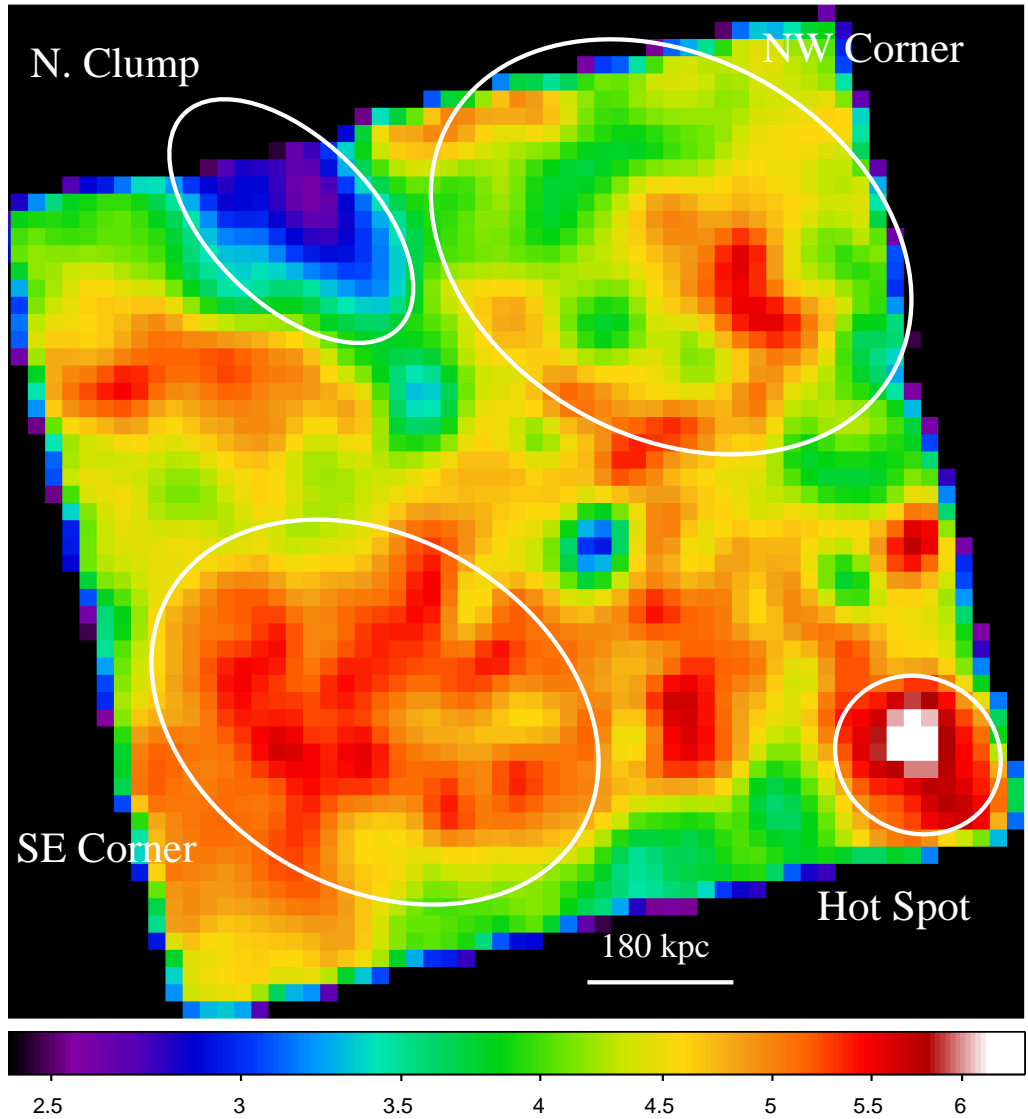}
\caption{Coarsely binned (20\arcsec~pix$^{-1}$) temperature map
  of the entire ACIS-I FOV for Obs-ID~9897, smoothed with a 2~pixel
  Gaussian. 
  The color-bar gives the temperature in keV.
\label{fig:acisi_tmap}
}
\end{figure}

\begin{figure}
\plotone{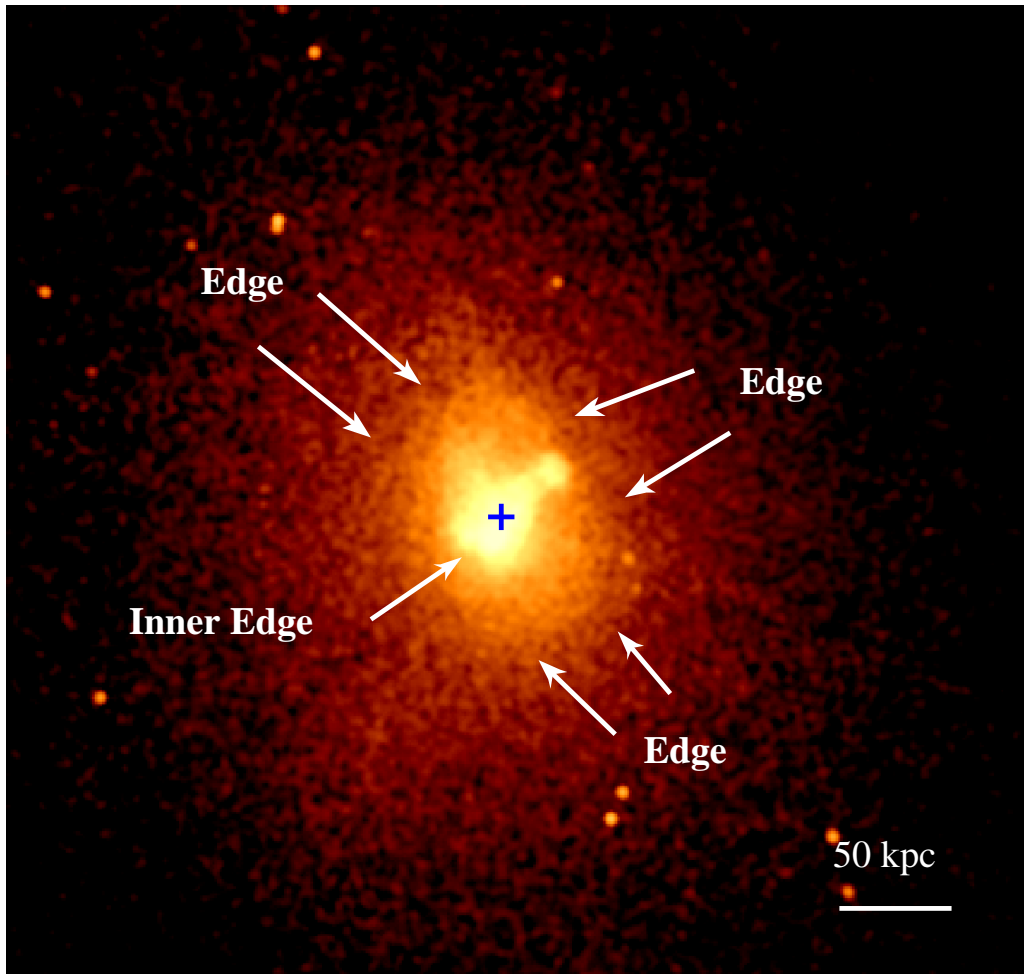}
\caption{ 0.3--2.0~keV corrected {\it Chandra} image, smoothed with a
  6\arcsec\ radius Gaussian.  In addition to the inner edge to the southeast,
  associated with the central core emission, the central core is
  encompassed by an outer elliptical edge, ranging in distance from
  about 30-50~kpc of our adopted central position (marked with a blue cross).
\label{fig:edges_img}
}
\end{figure}

\begin{figure}
\includegraphics[width=7in]{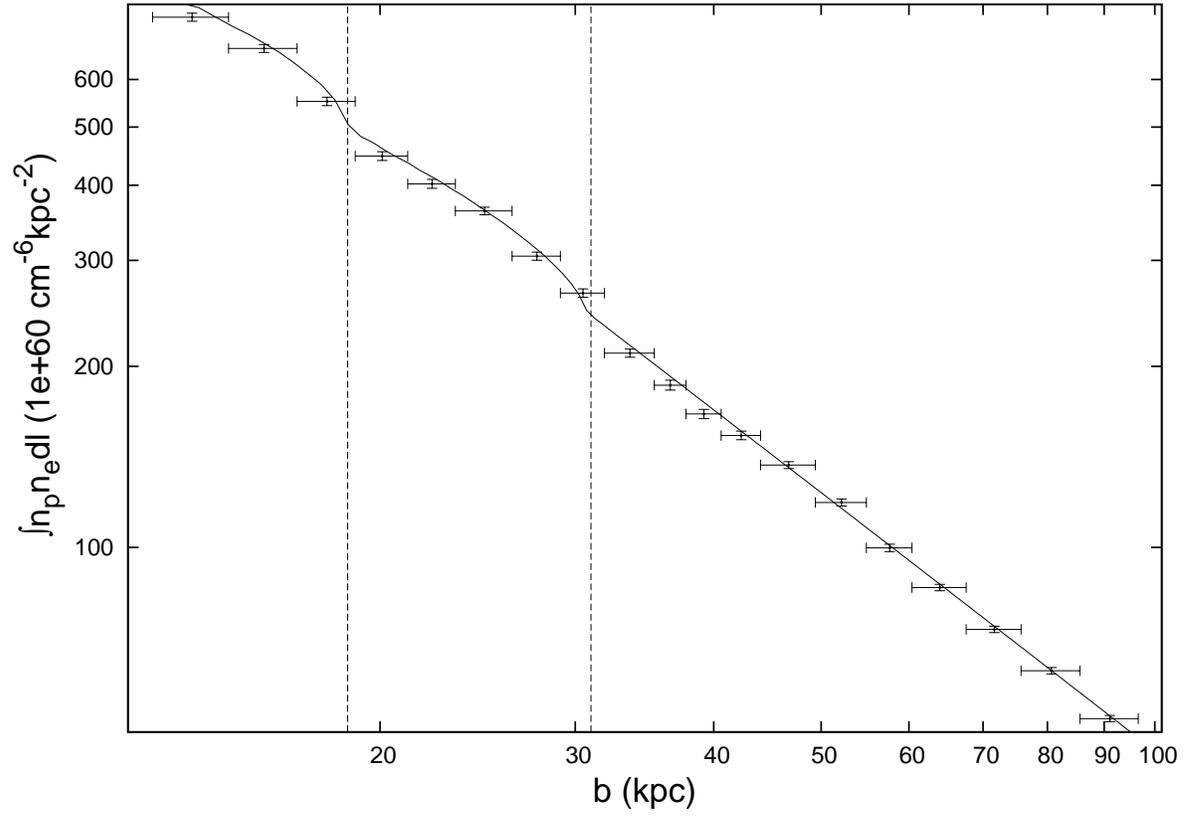}
\caption{
  Integrated emission measure profile, plotted against the
  semi-minor axis of the elliptical annuli and fit with a projected 3D density
  model (see text for details).  The solid line shows the model, and
  the vertical dashed lines mark the best-fitting break radii,
  corresponding to density discontinuities.
\label{fig:emprof}
}
\end{figure}

\begin{figure}
\plotone{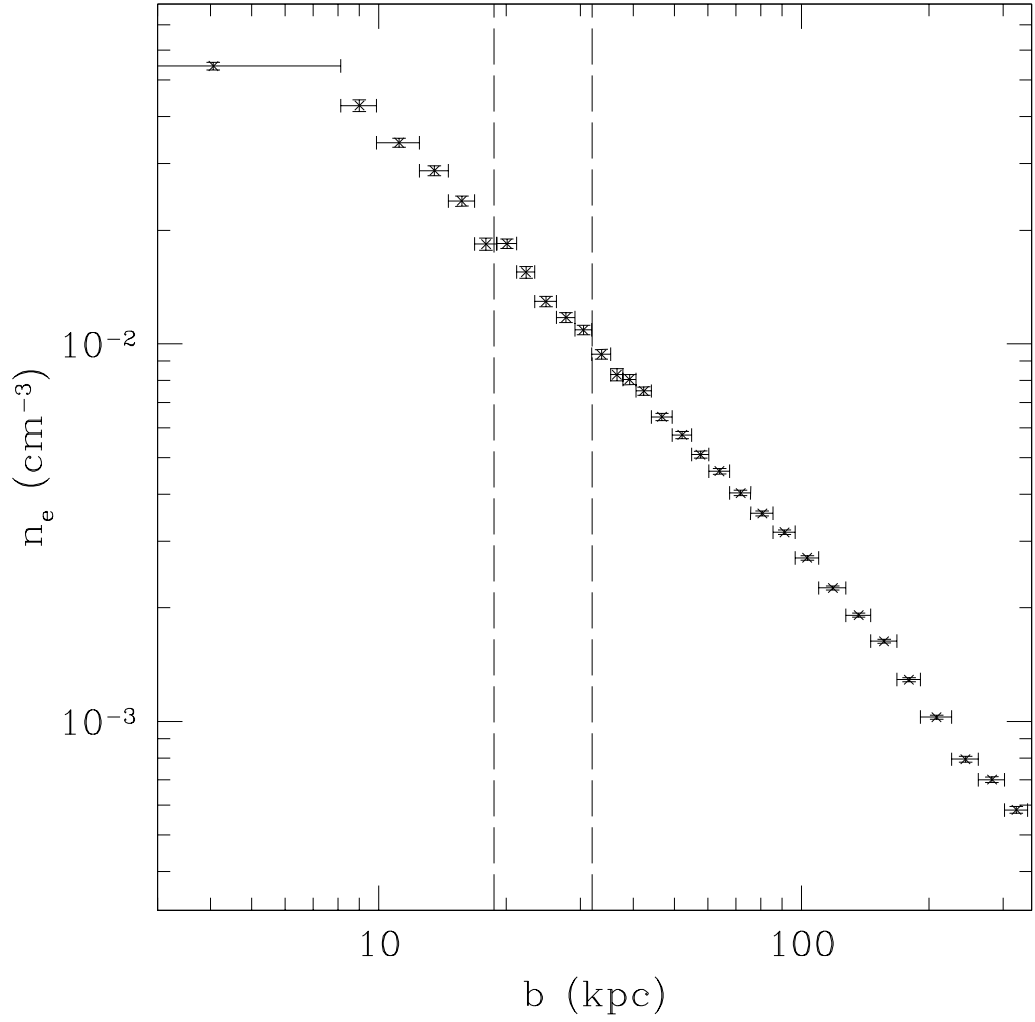}
\caption{Deprojected electron density profile plotted against the
  semi-minor axis of the elliptical annuli.  The dashed lines mark the
  best-fit break radii (see text).
\label{fig:neprof}
}
\end{figure}

\begin{figure}
\plotone{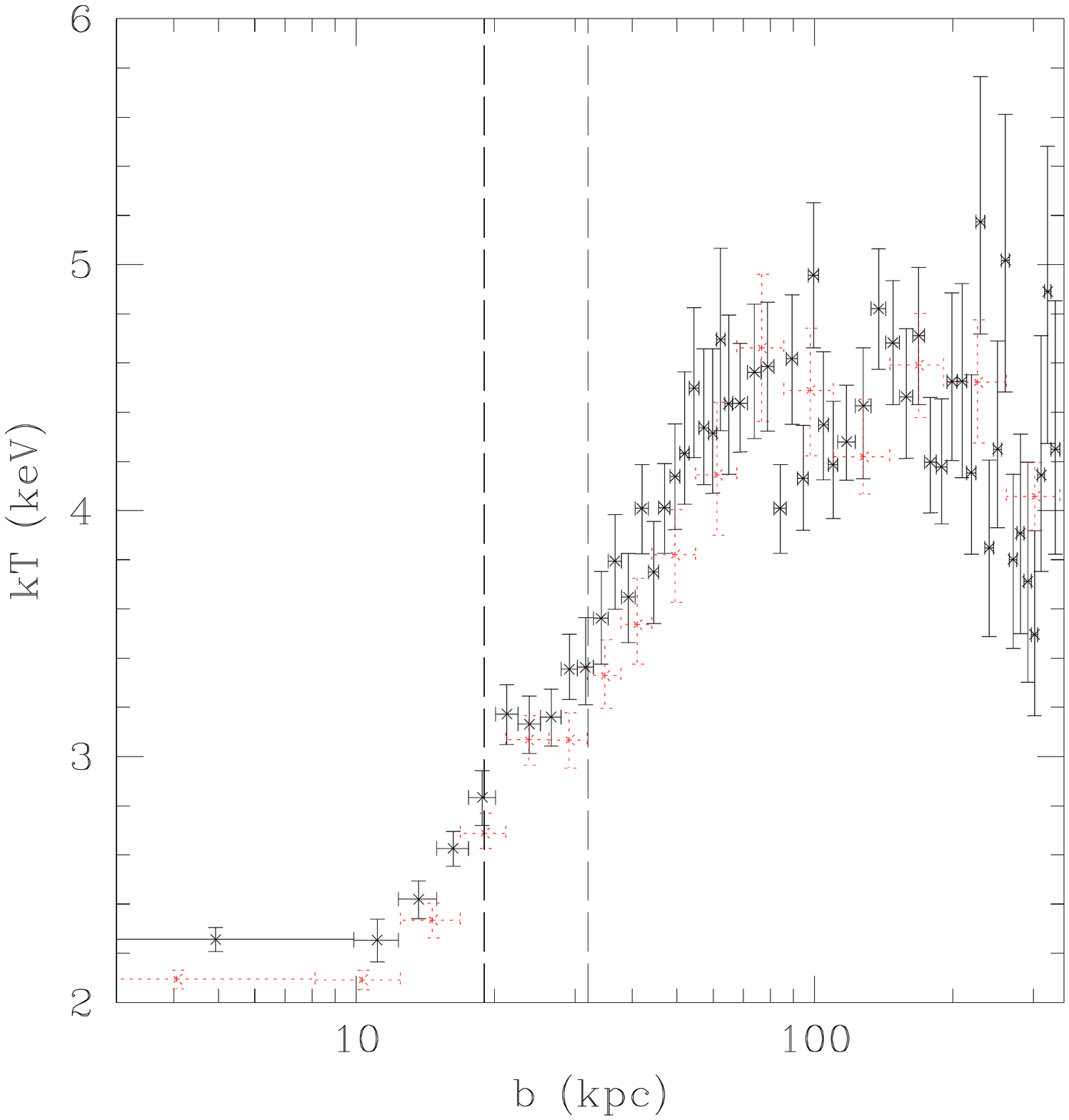}
\caption{Projected (solid black lines) and deprojected (dotted red
  lines) temperature profiles plotted against the
  semi-minor axis of the elliptical annuli.  The dashed lines mark the
  best-fit break radii (see text).
\label{fig:ktprof}
}
\end{figure}

\begin{figure}
\plotone{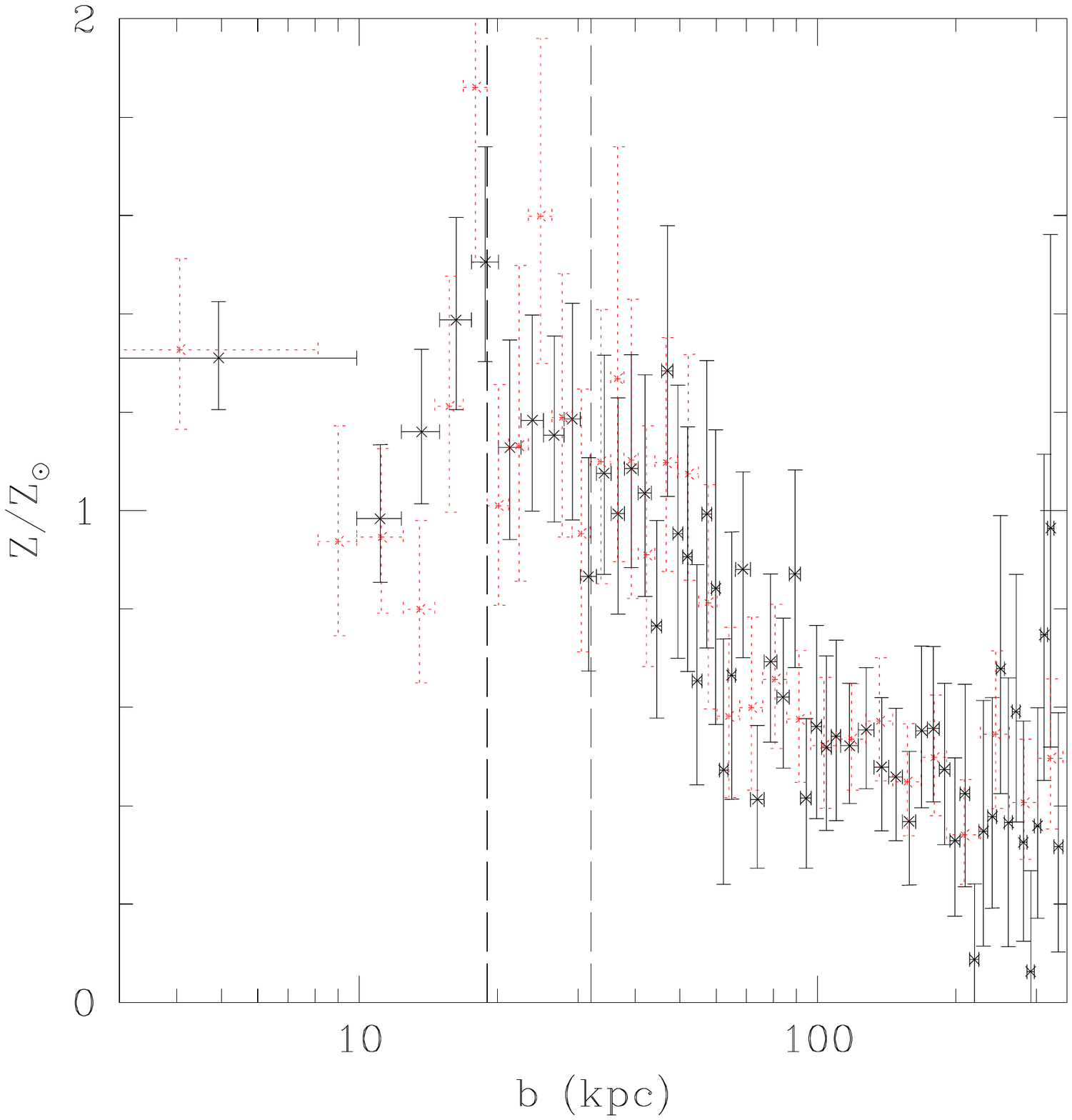}
\caption{Projected (solid black lines) and deprojected (dotted red
  lines) abundance profiles plotted against the
  semi-minor axis of the elliptical annuli.  The dashed lines mark the
  best-fit break radii (see text).
\label{fig:abprof}
}
\end{figure}

\begin{figure}
\plotone{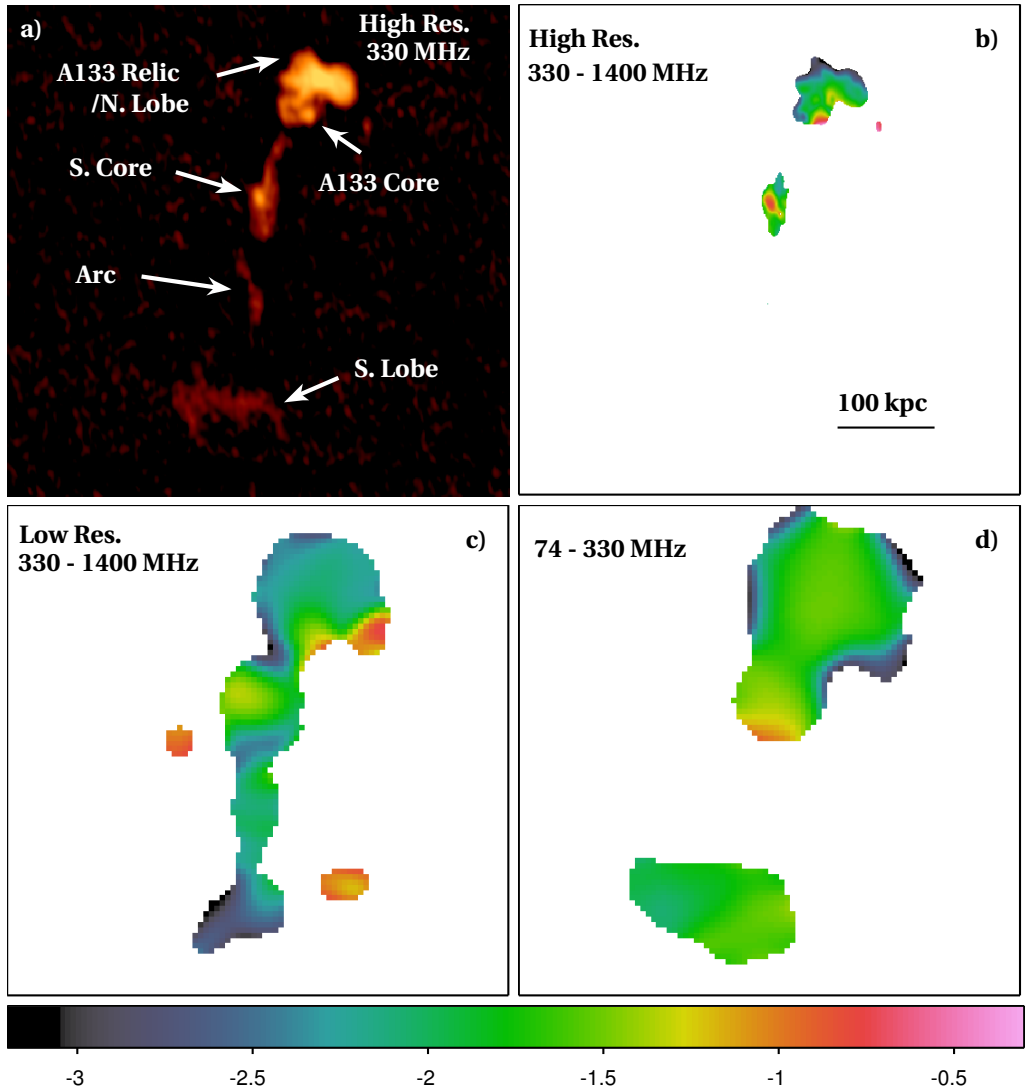}
\caption{Spectral index maps derived from radio observations at
  multiple frequencies. The color bar indicates the spectral index
  value in the upper-right and two lower panels.
  {\it a)} High-resolution 330~MHz A
  configuration radio map, as shown in Figure~\ref{fig:rad_imgs}c, for
  comparison.
  {\it b)} Spectral index map constructed from the higher resolution
  330~MHz and 1400~MHz observations.
  {\it c)} Same, but from the lower resolution 330~MHz and 1400~MHz
  observations.
  {\it d)} Low frequency index map, constructed from the 74~MHz and lower
  resolution 330~MHz observations.
\label{fig:index_maps}
}
\end{figure}

\begin{figure}
\includegraphics[angle=90]{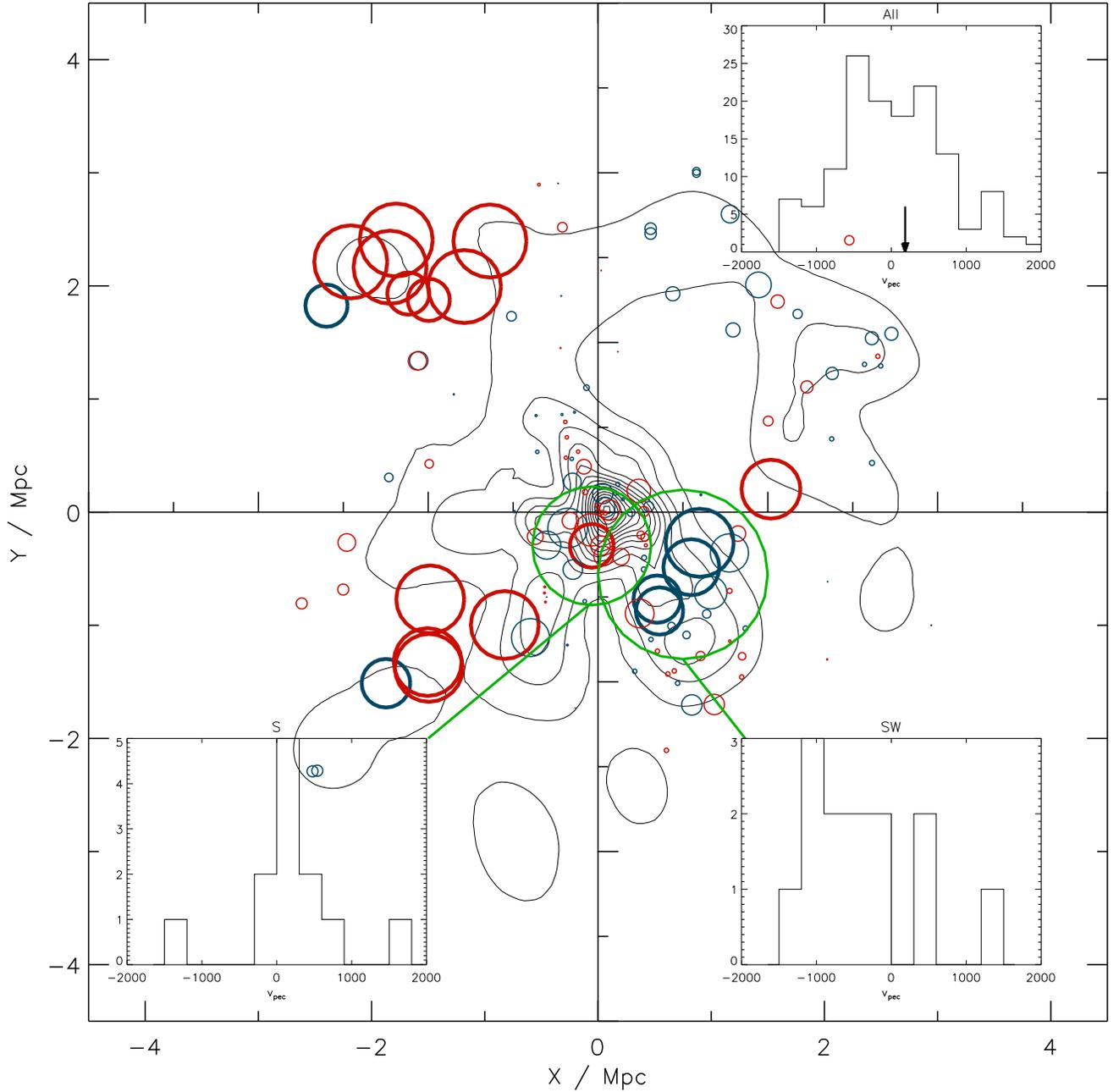}
\caption{Bubble plot for Abell~133.   The blue/red circles have
  peculiar velocities that 
  are negative/positive relative to the mean cluster redshift.  
  The circle radii are proportional
  to the log of the probability that the local velocity
  distribution differs from 
  the global one.   Bold
  circles indicate realizations that only occurred in 10\% or less of
  5000 Monte Carlo simulations where velocities were randomly shuffled
  between galaxies.  Overlaid are the galaxy density contours
  (smoothed with a variable-width Gaussian), where
  cluster members were selected from the {\it SuperCOSMOS} survey by
  requiring that they lie within $\pm0.2$~mag of the red-sequence.
  The contour levels begin at 20~galaxies~Mpc$^{-2}$ and are spaced by
  10.
  The insets show the radial velocity histograms for all of the
  galaxies (upper right, with the cD galaxy indicated by the arrow),
  the substructure to the southwest (lower right), and the
  substructure to the south (lower left).  The green circles indicate
  the regions used for the substructure velocity plots.
\label{fig:bubble}
}
\end{figure}

F02 consider four possibilities for the formation of the X-ray plume
indicated in Figure~\ref{fig:xr_core}: (1) a cooling wake, where hot
ICM gas is attracted into the wake of the moving cD galaxy and cools,
(2) convective lift initiated by the displacement of core gas due to
ram pressure stripping by a merging subcluster, (3) disruption of the
cold front (which they detect to the southeast) due to Kelvin-Helmholtz
instabilities, and (4) gas lifted by a buoyant bubble, in this case
the radio relic in Abell~133.  Based on arguments presented in F02 and
F04, they conclude that the plume has most likely been formed due to
uplifted gas by a buoyant bubble.  Examples of cool gas plumes capped,
and presumably uplifted by, mushroom-cap shaped radio bubbles are seen
in other systems (\eg, M87, Forman \etal\ 2007; NGC~5813, Randall
\etal\ 2010) and reproduced in numerical
simulations (Churazov \etal\ 2001).
In this scenario, the gas in the plume was once co-spatial with the gas
in the core.  If this is the case, we would expect the entropy in the
core and in the tip of the plume to be similar, assuming that the
plume gas has not conducted heat from or become mixed with the ambient ICM.
We therefore calculated the entropy for regions R1 and R3 (indicated in
Figure~\ref{fig:regs}) using results from the two-temperature fits
presented in Table~\ref{tab:xspec} and assuming a spherical geometry for
R3 and a prolate spheroidal geometry for R1, with the major axis in the
plane of the sky.  We find very similar entropy for each region, with
$S_{\rm R1} = 14.7^{+3.5}_{-4.1}$~keV~cm$^2$ and $S_{\rm R3} =
14.8^{+2.9}_{-6.0}$~keV~cm$^2$, supporting the idea that the gas has
been uplifted (however, not only are the statistical errors large, but
there are additional systematic errors from the assumptions about the
geometry, so this result is not conclusive).

As noted by F02, one concern is whether the bubble would be able to
uplift the mass of gas in the plume and remain buoyant.  Roughly, we
expect this to be the case if the mass of gas displaced by the bubble
is greater than the mass of entrained gas.  Assuming a spherical
geometry for the tip of the plume and a cylindrical geometry for the
plume (R3 and R2 in Figure~\ref{fig:regs}), we find densities and
masses in these regions of $n_{e, {\rm R3}} = 0.018$~cm$^{-3}$, 
$M_{\rm R3} = 6.8 \times 10^8 \msun$, $n_{e, {\rm R2}} =
0.0065$~cm$^{-3}$, and $M_{\rm R2} = 1.0 \times 10^9 \msun$.  The
total mass of the plume is therefore $M_{plume} = 1.7 \times 10^9
\msun$, which is consistent with the estimate given by F02
of $M_{plume} \approx 2.0 \times 10^9 \msun$.  Taking the average ICM
density at the location of the relic from Figure~\ref{fig:neprof}, and
assuming an oblate spheroidal geometry for the relic, with the minor
axis in the plane of the sky, we find that the mass of gas displaced
by the bubble is approximately 
$M_{displaced} \approx 1.1 \times 10^{10} \msun$, more than 6 times
the gas mass in the plume.  
As a check on this result, we did the same test with the eastern
filament and eastern ear-shaped radio lobe in M87 (Forman \etal\
2007).  In the case of M87, we find $M_{plume} \approx 1.6 \times 10^9
\msun$ and $M_{displaced} \approx 5.6 \times 10^{9} \msun$, so that
the displaced gas mass is roughly 3.5 times the mass in the plume, on
the order of what we see in Abell~133 where a very similar process
seems to be at work.
We conclude that, in principle, the bubble
is capable of buoyantly lifting the cool gas in the plume, and that this is
formation mechanism of the plume is fully consistent with the
observations and with the conclusions of F04.

Another possibility is that the plume was formed by the gas sloshing in
the core.  Consider Figure~7 in Markevitch \& Ascasibar (2006), which
shows the time evolution of merger induced gas sloshing.  At 1.9~Gyr,
a cool plume feature forms, which, if viewed from the right or left 
(within the merger plane) would look similar to the structure seen in
Abell~133.  In this interpretation, the cool spot at the tip of the
plume corresponds to the end of the spiral trail of the innermost cool
gas in the simulations, and is extended along the line of sight.
The radio relic could have formed at the core prior to the onset of
sloshing, and subsequently dragged along by the sloshing motions of the
gas, and thus appears to be at the tip of the plume.  
This requires that the perturber passed from the northwest to the southeast of
the cluster (in projection), on a trajectory perpendicular to that in
our model proposed to
explain the elliptical cold front edge.  In this scenario, the
southeastern subcluster reported by Dalton \etal\ (1997) provides a
possible perturber candidate.
Although this
model is plausible, it is difficult to test without detailed
hydrodynamic simulations, which are beyond the scope of this paper.

\subsection{The X-ray Wings} \label{sec:wings}

F04 argue that the formation of the X-ray wings, indicated in
Figure~\ref{fig:xr_core}, is due to a weak merger shock passing
through the cool core from the southeast and disrupting it.  Their
interpretation is based on a detection of a thermal gradient along the
plume such that it is cooler northwest of the boundary defined by the
wing features indicated in Figure~\ref{fig:tmap_core}a,d, and a
similar jump in temperature across the edge of the southwestern wing
(they do not detect a temperature difference across the northeastern
wing).  The edges of the wings therefore represent the edge of the
shock front, which has been distorted as it passed through the dense
core of Abell~133, giving it a curved appearance.  
The edge in the pseudo-pressure map north
of the cD galaxy, which coincides with the northern edge of the wing
features in the X-ray image (compare the overlaid curve in
Figures~\ref{fig:tmap_core}a~\&~\ref{fig:tmap_core}d), seems to
support this scenario. 
However, as F04
point out, the ``evidence is 
marginal because 
it is based on keeping the temperatures of the projected hot component
fixed, and the evidence disappears when all temperatures are left
free''.  With the new {\it Chandra} data, we are able to fit
two-temperature models in these regions without fixing temperature
parameters, thereby more accurately accounting for projection
effects.  As shown in Table~\ref{tab:xspec}, we find no evidence for a 
change in temperature between the core (R1), the plume (R2), and the
tip of the plume (R3).  In \S~\ref{sec:dxspec}, we did a further test
of this result by dividing the plume (R2) into 7 contiguous regions,
each with $\sim2000$ net counts, and fitting each region with a
two-temperature model with fixed
temperature and abundance for the projected hot component.  We find
that all of the fitted temperatures are consistent with 1.1~keV within
3$\sigma$, and most are consistent within 1$\sigma$.

Table~\ref{tab:wings} gives our results from spectral fits across the
wings, with regions defined as in F04.
The
single-temperature fits show no evidence for upstream and downstream
temperature differences, consistent with what F04 find.  The two
temperature fits show an {\it increase} in the lower temperature across the
northern wing, but not the western (this is in contrast to F04, who
find a {\it decrease} across the western wing, but not the northern).
However, for the
upstream regions the best-fitting higher temperatures are unreasonably 
hot and essentially unconstrained.  We interpret this as indicating 
that the upstream regions are simply devoid of cool gas from the core,
and the two temperature model instead mostly deprojects the inner
2-3~keV gas from 5-6~keV gas at larger radii (seen in the temperature
map in Figure~\ref{fig:tmap}), which is only poorly measured in these
fits.  
Alternatively, since the upstream regions are coincident with the radio
relic, the measurement of such high temperatures may indicate
the presence of very hot gas within the bubbles, or could be the
result of fitting a thermal model to a non-thermal component,
presumably arising from IC scattering of the CMB by
synchrotron-emitting particles within
the relic (see the discussion of region R4 in \S~\ref{sec:dxspec}).
Either interpretation is consistent with the upstream regions
corresponding to
X-ray cavities created by the radio relic pushing out the cool central
gas in these regions, as suggested by B\^{i}rzan \etal\ (2004; 2008),
creating the wing morphology seen in the surface brightness map.
This is also consistent with the detection of the edge in the
pseudo-pressure map along the northern edge of the X-ray wings since
the decrease in X-ray 
pressure in this region can be attributed to the additional support from the
radio-filled cavities.
Since we find cooler low-temperature gas downstream in the northern
wing we conclude that our results are inconsistent with the F04 model
of a shock passing through the core from the southeast, but are
consistent with the wing features arising from the displacement of
cool gas by the buoyantly rising radio-filled relic.  

\subsection{Radio Relic and Lobes} \label{sec:relic}

There have been at least two suggestions as to the origin of the
diffuse radio source in Abell~133 (which we refer to as the ``relic''
or ``northern lobe'', depending on context).  First, it may be a radio
relic, as suggested by its steep spectral index (it is the third
steepest source in the sample of 24
objects from B\^{i}rzan \etal\ 2008) and non-central
location (S01).  However, some of the properties of the source seem to be at
variance with radio relics observed in other clusters.  While we have
argued that Abell~133 is dynamically disturbed, there is no clear
indication of a merger shock, which is thought to be required to
re-accelerate the radio emitting particles (although see the discussion of
Figure~\ref{fig:acisi_tmap} in \S~\ref{sec:dynamics}).  Furthermore, the
integrated polarization is low ($\sim 2.3$\%; S01) and shows
complicated spatial variations (F02), in contrast to
most radio relics.  Second, this feature may be an old
detached radio lobe associated with the central source in the cD
galaxy (Rizza \etal\ 2000), similar to the attached eastern ``ear''
lobe in M87 (Owen
\etal\ 2000; Forman \etal\ 2007) and consistent with our above
conclusion that the X-ray plume was formed via buoyant lift by a
rising bubble.  However, in such sources one typically finds two lobes
produced by a double-sided jet (although there are examples of
single-lobed systems, \eg, Bagchi \etal\ 2009).
In this section we re-examine the nature of this source in the context
of our conclusions about the dynamical state of Abell~133 and the
large-scale double-lobed morphology revealed by deeper radio observations
(Figure~\ref{fig:rad_imgs}).

\subsubsection{Relic Rise Time, Synchrotron Age, and Morphology} \label{sec:rise}

Since we have concluded that the X-ray plume was formed due to buoyant
lift of cool gas by the rising relic, it is of interest to compare
the relic age to the buoyant rise time.
We can estimate a lower-limit on rise-time of the relic by assuming
that the bubble rises with an average speed equal to the sound speed
in the gas.  For a local ambient ICM temperature of 2.3~keV
(consistent with results in Table~\ref{tab:xspec}) the sound speed is
782~\kms, giving a lower-limit rise-time of $t_{rise} > 4.6 \times
10^7$~yr.  
This is a lower limit
since the bubble need not rise in the plane of the sky, and since the
bubble may rise at some fraction $f$ of the sound speed.  If we take
average values and assume that 
$f \approx 0.5$ (as suggested by Churazov \etal\ 2001), and that the 
bubble rises along a vector 45\mydeg\ from the plane of the sky, the
rise time increases to $t_{rise} \approx 1.4 \times 10^8$~yr.
B\^{i}rzan \etal\ (2008) estimate the synchrotron age of the radio relic to be
$t_{relic} > 4.8 \times 10^7$~yr.  
Thus, the lower limits placed on $t_{rise}$ and
$t_{relic}$ allow $t_{rise} < t_{relic}$, as expected (\ie, the age of
the relic should be more than the time it took to rise to its current
position).
However, using low frequency radio observations, particularly at
29.9~MHz (Finlay \& Jones 1973) and 80~MHz (Slee 1995), S01 estimate a
synchrotron age of $t_{relic} = 4.9 \times 10^7$~yr, 
close to our estimated lower limit on $t_{rise}$, but shorter than the
``average'' rise time of $t_{rise} \approx 1.4 \times 10^8$~yr given above.
Based on their value of $t_{relic}$ and the assumption that the relic
likely does not rise in the plane of the sky, S01 conclude that the
rise time and the age of the relic are inconsistent, and suggest that
the relic may 
have come from the galaxy directly west of the core in Abell~133
(visible, e.g., in Figure~\ref{fig:rad_imgs}a) rather than from the cD
galaxy itself.  

More recent {\it Chandra}
observations make it clear that the relic has in fact risen from the
central cD galaxy, inconsistent with the S01's suggestion that the
relic has come from another galaxy just outside of the core.
We note that the estimate of the relic age by S01 may be inaccurate,
both because the low frequency radio observations have very poor
resolution ($\sim1$\mydeg) and fluxes must be corrected by subtracting
off the contribution from other sources in the field (which requires
some assumption about spectral indices), and because the flux in the
region of the relic may be
contaminated by the 
putative background lobe discussed in \S~\ref{sec:lobes} (although an
estimate of the latter effect is difficult given the data we consider here).
As a check on the relic age estimate of S01, we used the model
applied, \eg, in Parma \etal\ (2007) and Giacintucci \etal\ (2007),
including the new radio data to estimate the synchrotron age.  Our 
results are fully consistent with those of S01.  We conclude that an
improved estimate of the relic age will require additional low
frequency radio data below 74~MHz (\eg, from {\it LOFAR}).

If we assume that the relic age given by
S01 is accurate, the most natural solution to the inconsistency noted
by S01 is that
the bubble rises along a vector that is close to the plane of the sky,
so that the rise time is close to the lower limit and consistent with
$t_{relic} = 4.9 \times 10^7$~yr.  That this is the case is suggested
by the X-ray images alone, specifically by the fact that we see cavities in the
cooler central ICM that are co-spatial with the relic (detailed spectral
fits indicate that the cavities are devoid of cooler
$\sim1.5$~keV gas, see \S~\ref{sec:dxspec}).  If the relic was much in
front of or behind the core, such that it was beyond the cooler
gas encompassed by the edge indicated in Figure~\ref{fig:edges_img}, we
wouldn't expect to see cavities in the cool gas (in fact, we would not
expect to see cavities at all, since the lower surface brightnesses at
larger radii make density contrasts more difficult to detect [B\^{i}rzan
\etal\ 2009]). If the central gas were significantly extended along
the line of sight the bubble may have risen along this extension,
although not only would it still be difficult to detect the bubble for
large inclination angles of the extension due to the large column
density of cool gas behind (or in front of) the bubble, but, as we
show in \S~\ref{sec:profs}, the surface brightness profile is
well-described by an ellipsoidal density model with the major axis in
the plane of the sky.
Thus, we conclude that the relic has likely risen in a direction that is
close to the plane of the sky, resolving the inconsistency between
$t_{rise}$ and $t_{relic}$ reported by S01.

However, we note that our interpretation of Abell~133
as a dynamically disturbed system can in principle resolve a
situation where $t_{rise} > t_{relic}$, and it is of general interest
to consider these effects.  
First, the cosmic ray particles in radio
relics are thought to be re-energized by compression from merger shocks,
as proposed by En\ss lin \& Gopal-Krishna (2001).  If the particles
have been re-accelerated due to a merger with a subgroup, the true age
of the relic will be larger 
than the synchrotron age estimated from the current radio luminosity
and spectral index (however, in \S~\ref{sec:dynamics} we showed that
for the potential perturbing subclusters
the time since the merger is $t_{merge} > 1.0 \times 10^8$~yr, also
longer than the S01 estimate of $t_{relic}$).  
Second, we note that the close passage of a
perturbing subgroup, which has initiated gas sloshing in the core,
can in principle decrease the rise time required to get to the current position.
The passage will temporarily deepen the gravitational potential,
causing gas to flow towards the subgroup and increasing the
buoyant force on the bubble.  To estimate the
strength of this effect, we used the mass model for Abell~133
described in \S~\ref{sec:dynamics} to estimate the duration and speed
of the core passage phase of a putative subcluster.  
To maximize the effect, we assumed an impact parameter of 37~kpc
(equal to the projected distance between the relic and the AGN) and
that the subcluster mass
profile at small radii is the same as that of Abell~133, truncated at
100~kpc. (Although the 
subcluster's total mass will be less than that of the main cluster it
may be more concentrated, giving more mass in the core.  However the
core of the subcluster cannot be 
more massive than the core of the main cluster otherwise the main
core would have been completely destroyed by the merger, so we
maximize the effect by 
making the core and subcluster masses equal.)  We find an average speed at core
passage of $v_{merge} \approx 2600$~\kms.
The buoyant acceleration of the bubble is
\begin{equation}\label{eq:bacc}
a_b = g \frac{\rho_a - \rho_b}{\rho_a + \rho_b},
\end{equation}
where $\rho_b$ is the density in the bubble, $\rho_a$ is the
density of the ambient ICM, and $g$ is the gravitational
acceleration.  We assume that $\rho_a \gg \rho_b$, so that $a_b \approx
g$.  The maximum velocity the bubble can attain is determined by the
drag force, which is given by
\begin{equation}\label{eq:drag}
F_d \sim \frac{1}{2}C v^2 \rho_a A,
\end{equation}
where $A$ is the cross-sectional area of the bubble and $C$ is the
drag coefficient (we use $C = 0.75$, see Churazov \etal\ 2001).
Therefore, balancing the buoyant force $F_b = (\rho_a + \rho_b) a_b$
with $F_d$ gives a maximum bubble velocity of
\begin{equation}\label{vbub}
v_{max} = \sqrt{g \frac{8 r}{3 C} \frac{\rho_a - \rho_b}{\rho_a}}
\approx \sqrt{g \frac{8 r}{3 C}},
\end{equation}
where $r \approx 10$~kpc is the semi-minor axis of the bubble.
Using the above model and starting the bubble at rest 5~kpc from the
core gives $t_{rise} = 9.1 \times 10^7$~yr neglecting the contribution
from the subcluster, and $t_{rise} = 7.6 \times 10^7$~yr including the
subcluster.  Thus, even a favorable estimate of the subcluster's
influence on the rise time gives gives only a 16\% effect,
insufficient to reduce the average rise time enough to be consistent
with the estimate of $t_{relic}$ from S01.

Our proposed merger/gas sloshing model can have some additional
consequences for the radio observations, in
particular when considering the morphology of the relic and jet.
As noted by S01, the 1.4~GHz image (Figure~\ref{fig:rad_imgs}a) shows a 
clear separation between the loop feature extending east of the
Abell~133 core and the filamentary radio relic.  However, at lower
frequencies the jet and relic seem to connect
(Figure~\ref{fig:rad_imgs}c), and Rizza \etal\ (2000) report a
connection in their 1.4~GHz observations.  
If we assume that features are indeed connected, one then needs to
explain not only the apparent misalignment between the inner jet
(which points east by northeast) and the direction of the bubble from
the core (north by northwest), but also the lack of a counter jet and
relic as observed in other systems (\eg, M87).
We note that not only can gas sloshing, in
principle, push around the lighter radio plasma (without affecting the
cooler, denser, X-ray plume), deforming the jet and
giving rise to some of the fine filaments seen in the high-resolution
radio images, but the additional buoyant force from a passing
subcluster can change the rise trajectory of buoyant bubbles, possibly
pushing two bubbles together as they rise from the core (such ``misalignment''
of buoyantly rising bubbles is clearly observed in other systems, \eg,
M87 [Forman \etal\ 2007]).
To test this effect, we applied the same model we used to
calculate the effect of the subcluster on the buoyant rise time
(described above).  We used a smaller velocity at core passage of
1000~\kms and evolved the bubbles starting just prior to core passage to
maximize the effect.

\begin{figure}
\plotone{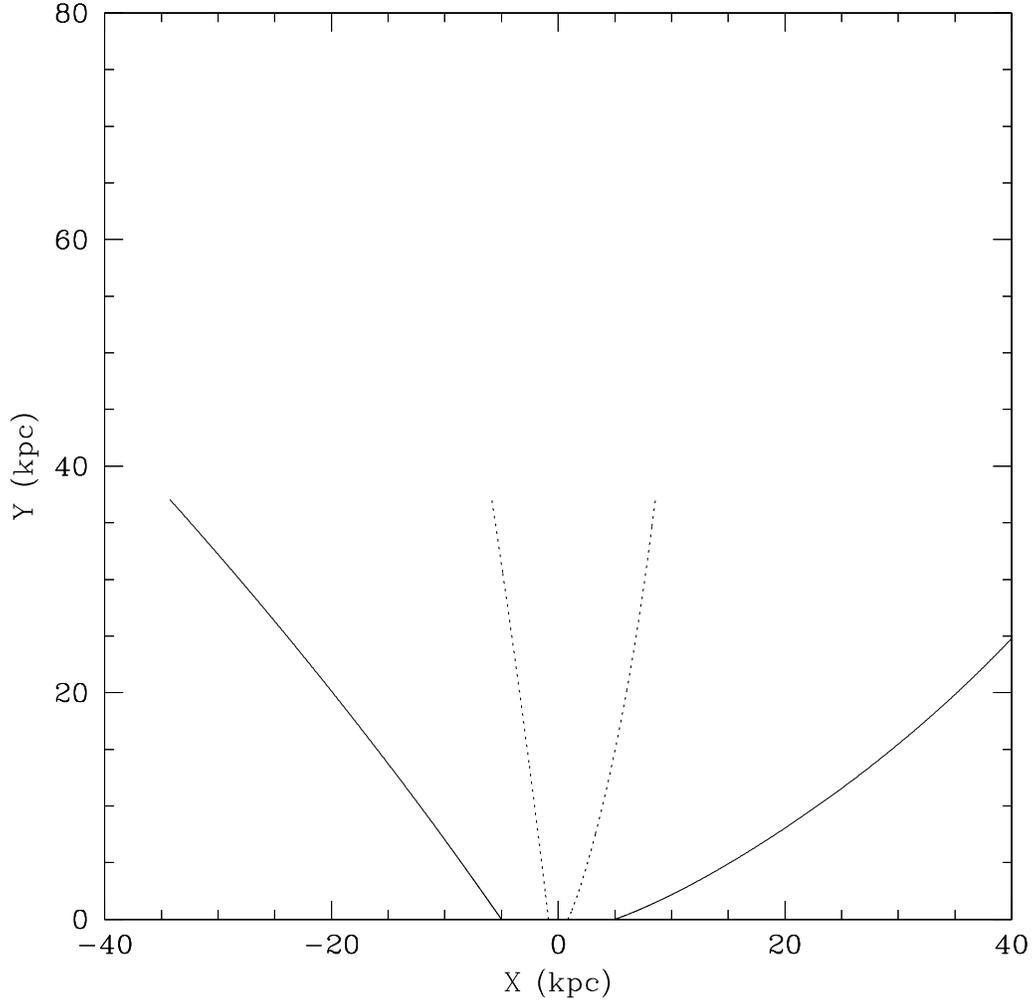}
\caption{
The rise trajectories of two buoyant radio bubbles, including the
effects of a putative subcluster that passes the cluster core (at (0,0))
along the line $y = -37$.  Initially, the bubbles are 5~kpc from the
central AGN and are released just prior to core passage.  If the
effects of the subcluster are removed, the bubbles rise along the
$x$-axis.  Solid lines indicate the side view, where the subcluster
and bubbles move in the $x-y$ plane, and dotted lines indicate the
trajectories as viewed after an 80\mydeg\ rotation about the
$y$-axis.  The rise time is $1 \times 10^8$~yr.
\label{fig:2d_bubbles}
}
\end{figure}

First, we consider the case where the subcluster moves in the same
plane as the jets.  The resulting trajectories of the bubbles are shown
in Figure~\ref{fig:2d_bubbles}.  In this figure, the subcluster passes
along the line $y = -37$, moving from left to right.  Without the
influence of the subcluster, the bubbles rise buoyantly along the
$x$-axis.  We see that while the subcluster has a significant effect on the
rise trajectories, the final separation between the bubbles is too
large for them to overlap.  However, if we view this case at an angle
such that the initial direction of the jets (and the direction of
motion of the subcluster) is close to our line of sight (consistent
with our suggestion that the motion of the perturbing subcluster
is not in the plane of the sky), then the
bubbles can overlap in projection (see the dotted lines in
Figure~\ref{fig:2d_bubbles}).  In this projection, the jets still
point along the $x$-axis, giving rise to a misalignment between the
jet direction and the bubble trajectory, similar to what is observed
(Figure~\ref{fig:rad_imgs}a).  
Therefore, the additional buoyant force of a merging subcluster can,
in principle, address the primary objection to the interpretation
of the relic as a detached radio lobe (namely that only one lobe is
observed), although there are some difficulties with the details of our model.
In this simulation, the rise time of the
bubbles is $\sim 1 \times 10^8$~yr, and the final true separation between
the bubbles and the core is $\sim 50$~kpc.  The separation places the
bubbles just outside the central cool gas, although, as we argue above,
the X-ray images suggest that the relic has evacuated cavities in the
cool central gas, and therefore cannot be much more than $\sim 40$~kpc
from the core and must have risen close to the plane of the sky (since
the projected separation from the core is $\sim 37$~kpc).  Thus, while
this model can potentially reproduce some of the observed radio
features, it is not fully consistent with the observations (although we
note that sloshing induced gas motions can in principle add to the
rising speed of the bubble and resolve this inconsistency).

\begin{figure}
\plottwo{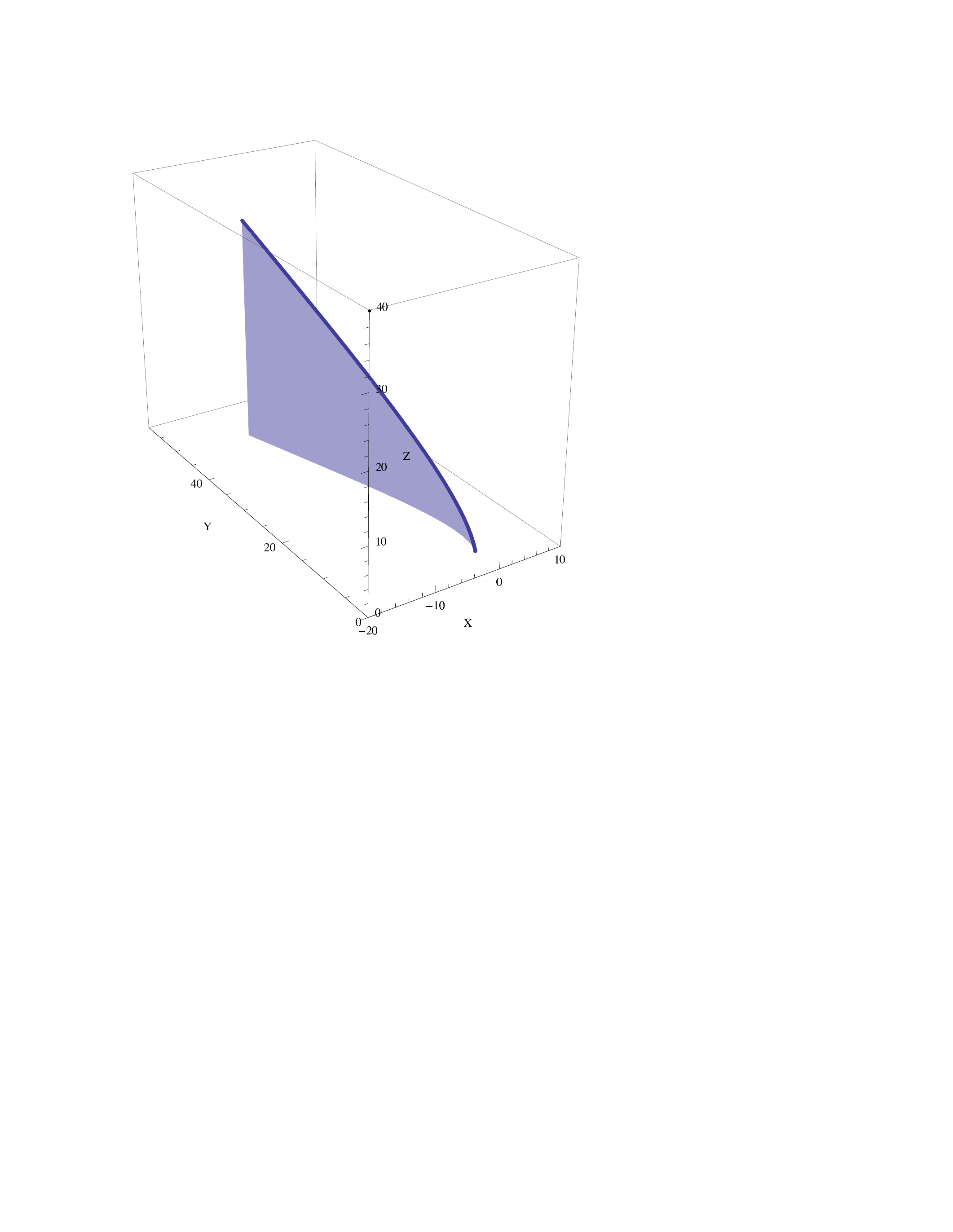}{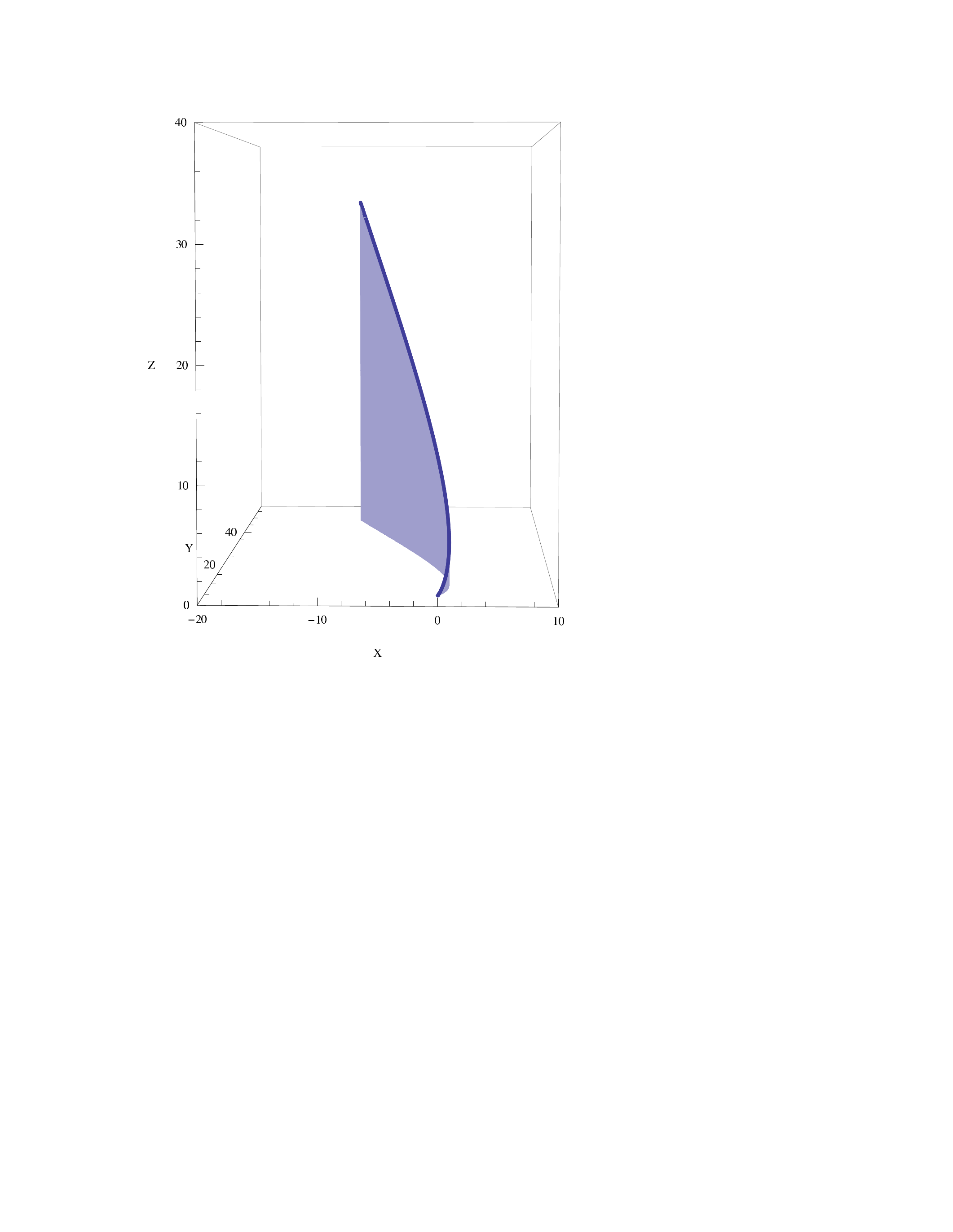}
\caption{
  Two views of a rising bubble trajectory in 3D, similar to
  Figure~\ref{fig:2d_bubbles} but with the subcluster passing
  perpendicular to the plane defined by the inner jets, along the line
  $z = -37$ (from $x<0$ to $x>0$).  The jet lies along $x = 0$. If the
  effects of the subcluster are removed, the bubble rises along
  $x=0$. The rise time is $1.2 \times 10^8$~yr.
  \label{fig:3d_bubbles}
}
\end{figure}

We also considered the case where the merger axis of the subcluster is
perpendicular to the plane defined by the inner jets.  The subcluster
passes by the core, equidistant from each bubble, so the effect on
each bubble is the same and the trajectories are symmetric.  The
resulting trajectory for a rising bubble is shown in
Figure~\ref{fig:3d_bubbles}.  The results are similar to the case
where the subcluster passes parallel to the inner jet axis: while the
additional buoyant force from the subcluster can significantly affect
the rise trajectories and give a possible projection where the bubbles
are co-spatial and separated from the central AGN, the final distance
from the core ($\sim60$~kpc) puts the bubbles outside of the central
cool gas and is inconsistent with the observation of cavities in the gas.
We conclude that, although we have shown that an interloping subcluster
can in principle have a significant effect on the morphology and
evolution of buoyant radio bubbles, our simplistic model fails to
reproduce what we see in Abell~133 (although the added effect of
sloshing-induced gas motions could resolve this discrepancy).

\subsubsection{Radio Lobes} \label{sec:lobes}

The overall structure of the low-resolution, low-frequency radio
images (Figure~\ref{fig:rad_imgs}) and spectral index maps
(Figure~\ref{fig:index_maps}) is consistent with what one would expect for
a double-lobed radio galaxy.  Optical observations place the galaxy core at a
redshift of $z=0.293$, much larger than the redshift of Abell~133
($z=0.057$).  However, the high-resolution 1.4~GHz radio image reveals the
bright, complex relic in the region of the northern lobe, and X-ray
observations unambiguously associate the relic with cool central gas
in Abell~133.  We are therefore left with three possibilities: (1) the
radio galaxy is associated with Abell~133, and the radio core overlaps
with the optical source by chance, as does the northern lobe with the
relic; (2) the core and jet features are at $z=0.293$, but the lobes
are unassociated with this source (with at least the northern
lobe/relic being in Abell~133), and the alignment between the jets/arc
and the lobes is
coincidental; and (3) there is a giant background radio galaxy,
oriented such that the northern lobe overlies the relic in
Abell~133.  Unfortunately, we are unable to distinguish between these
three scenarios with the current data (although we searched for a
surface brightness depression associated with the southern lobe and
found none, the low X-ray surface brightness at this radius would make
any putative cavity associated with the lobe extremely difficult to
detect [B\^{i}rzan \etal\ 2009], so the non-detection cannot be used
to argue that the southern lobe is a background source).  
Out of these three possibilities, we suggest that the third option of
a double-lobed background radio galaxy is most likely since it requires the
smallest number of random coincidences.

Assuming that the radio observations represent a radio
galaxy at $z = 0.293$, we find that the total extent of the lobes is
1.6~Mpc (compared to 400~kpc if the galaxy were at the
redshift of Abell~133).  This is within the range of what has been
found for other giant radio galaxies (Schoenmakers \etal\ 2001 find a
strong drop in the number of sources larger than $\sim2$~Mpc).
We do not measure IC emission in the region of the southern lobe, and
therefore cannot infer the energy density in electrons ($u_{\rm e}$)
and magnetic fields ($u_{\rm m}$), but the size of the radio galaxy
(which correlates linearly with these properties, Isobe \etal\ 2009)
implies $u_{\rm e} \approx 2 \times 10^{-13}$~ergs~cm$^{-3}$ and
$u_{\rm m} \approx 2 \times 10^{-14}$~ergs~cm$^{-3}$.

Flux contamination from a putative background lobe may explain some of
the properties of the relic in Abell~133 that seem to be at variance
with relics observed in other clusters.  
Blending with a background source may well lead to the complex polarization map
presented by F02.  Additionally, since the background lobe is expected
to be old (more than $10^8$~yrs) and to not contribute much flux
at high frequencies (as is typical for giant radio galaxy lobes) it
may steepen the combined spectrum.
To estimate the effects of the background lobe, we assume that it has
the same flux as the southern lobe.
(Note that this assumption need not be valid, since opposite
radio lobes can vary significantly in brightness.  An example of an
extreme case is the single-lobed giant radio galaxy reported by Bagchi
\etal\ [2009].)  We find that the background lobe contributes 13\%,
11\%, 7\%, and 3\% of the total flux in the region of the relic in the
74~MHz, high-resolution 330~MHz, low-resolution 330~MHz, and
low-resolution 1.4~GHz images respectively.  These flatten the
spectral indices of the relic (Table~\ref{tab:sindex}) by 2.2\% for
$\alpha_{330,{\rm A}-1400,{\rm B}}$, 2.9\% for $\alpha_{330,{\rm
    B}-1400,{\rm C}}$, and 0.7\% for $\alpha_{74,{\rm A}-330,{\rm
    B}}$, in each case within or on the order of the 1$\sigma$ error
range.  We conclude that the background northern lobe will not
significantly alter the spectral index for the relic unless it is
significantly brighter than the southern lobe.

In conclusion, we find it most likely that the large-scale
double-lobed feature represents a background giant radio galaxy, with
the northern lobe obscured by the brighter relic in Abell~133.
The nature of the radio relic remains unclear.  
Its morphology and association with a (presumably uplifted) plume of
cool gas is very similar to the eastern ``ear'' radio lobe feature in
M87 (Forman \etal\ 2007).
The putative
background northern lobe does not significantly flatten the spectral
index of the relic, and it is unclear whether it can explain the
unusual polarization map presented in F02.  The steep index suggests a
radio relic classification, and while we do not find clear
evidence of a merger shock, which would be required to re-accelerate
the particles in the relic, we have argued that Abell~133 is a
dynamically disturbed system, and find some evidence for shock heated
gas south of the cluster core.  
The free-fall time of the gas at the tip of the X-ray plume is
$\sim10^8$~yr, a factor of two larger than the most likely rise time
of the bubble of $t_{rise} \approx 4.6 \times 10^7$~yr, so that the
plume may persist for some time after it has decoupled from the
buoyant action of the relic.
In the case of the double-lobed radio
galaxy interpretation, we have shown that the passage
of a merging subcluster can in principle move the bubbles buoyantly so
that they overlap in projection and give something similar to what is
observed, although in our simple model the bubbles end up too far from
the core to produce the observed radio cavities.  We conclude that the
most likely interpretation is that the diffuse source is a relatively
young radio relic, re-energized by the dynamical activity in
Abell~133.  Its youth not only explains its proximity to the core (and
its association with the X-ray plume), but it would require less
energy to re-accelerate the younger particle population, in which case
a weak shock or even gas sloshing may provide the necessary energy to
compress the relic.  

%

\section{Summary} \label{sec:summary}

We have presented results from deep {\it Chandra} and {\it VLA}
observations of the cool core cluster Abell~133.  The elliptical X-ray
isophotes, the offset between the cool gas and cD galaxy, optical
substructure detections, and a weak elliptical sloshing edge (which is
sharpest to the southeast) suggest
that Abell~133 is dynamically disturbed.  The detection of multiple
peaks in the cD galaxy at optical, NIR, and UV
wavelengths and corresponding minima in the X-ray temperature map suggests an 
ongoing galaxy-galaxy merger, and the offset between the cool gas minima
and the peaks at longer wavelengths suggest bulk gas motion in the
core, consistent with gas sloshing from a recent subcluster merger.
The large scale temperature map reveals a cool clump of gas to the
north, possibly indicating stripped gas from a merging subcluster, and
somewhat higher temperature gas to the south (as compared to the
north), possibly due to ICM gas heating by a merger shock. Although
optical tests for substructure are inconclusive, there is evidence for
substructure to the south and southwest of the core.  We suggest a
model where a merging subcluster (possibly one of the optically
detected substructures) has passed southeast of the core,
leaving the cool clump of stripped gas to the north, shock heating
the ICM in the south, and initiating gas sloshing in the core.  Since
the subcluster does not move in the plane of the sky, we view the
sloshing spiral (seen in simulations and observations of other
systems) at an angle, resulting in the ragged elliptical surface
brightness edge we identify as a cold front.

We find that the bird-like morphology of the central gas can be fully
explained by the presence of the diffuse radio source in Abell~133.
The plume of cool gas is buoyantly lifted by the relic, as seen in
simulations and observations of other systems, and in agreement with
previous work.  Our results regarding the X-ray wings are inconsistent
with the previously suggested model where they arise from a shock
passing through the cool core since we rule out temperature jumps
across the putative shock front.  Instead, the wings are created by
X-ray cavities occupied by the radio relic on either side of the plume.

In the radio, we identify a previously unreported $1.6$~Mpc background giant
radio galaxy at $z = 0.293$.  The northern lobe of this source
overlaps with the relic in Abell~133, contributing modestly to the
total flux in this region and marginally affecting measured spectral
indices.  The nature of the steep-spectrum relic source in Abell~133
remains unclear.  
The agreement between the lower limit on the rise time
and synchrotron age of the relic, along with its association with X-ray
cavities in the central cool gas, suggest that the relic has risen
along a vector close to the plane of the sky, forming the cool X-ray
plume via buoyant lift.
X-ray spectral fits in the region of the relic indicate the presence of
either high-temperature gas or non-thermal emission.  Although the
measured photon index is flatter than the spectral slope inferred
from radio observations, indicating that the non-thermal emission is
not from IC scattering of the CMB by the radio emitting particles, the
slopes are consistent within the (large) 90\% confidence intervals.
We have
argued that it is most likely a young relic, possibly re-energized by
a weak shock or gas sloshing associated with the recent merger, although
other interpretations remain plausible.  We have shown that, in
principle, the additional buoyant force of a passing subcluster can
significantly affect the trajectories of buoyantly rising radio
bubbles and explain the drastic misalignment of the eastern inner jet
and the rise direction of the relic, although the details of our simple
model do not fully agree with the observations in the case of Abell~133.

\acknowledgments
The financial support for this
work was partially provided for by the Chandra X-ray Center through
NASA contract NAS8-03060, and the Smithsonian Institution.
The National Radio Astronomy Observatory (NRAO) is a facility of the
National Science Foundation operated under cooperative agreement by
Associated Universities, Inc. Basic research in radio astronomy at the
NRL is supported by 6.1 Base funding.  We thank Matteo Murgia for
access to private code for calculating synchrotron ages.
We thank E. L. Blanton and the anonymous referee for useful comments.

\clearpage

\begin{deluxetable}{lcccc}
\tablewidth{0pt}
\tablecaption{{\it Chandra} X-ray Observations of Abell~133 \label{tab:xray}}
\tablehead{
\colhead{Obs ID}&
\colhead{Date Obs}&
\colhead{Target}&
\colhead{CCDs Used}&
\colhead{Cleaned Exposure}\\
\colhead{}&
\colhead{}&
\colhead{}&
\colhead{}&
\colhead{(ksec)}
}
\startdata
2203&2000-10-13&A133 &S3, S1&30.0\\
3183&2002-06-24&A133-OFFSET &I0, I1, I2, I3, S2&43.0\\
3710&2002-06-26&A133-OFFSET&I0, I1, I2, I3, S2&42.0\\
9897&2008-08-29&A133 &I0, I1, I2, I3, S2&68.7\\
\enddata
\end{deluxetable}

\begin{deluxetable}{lcccccc}
\tablecaption{Radio Observations of Abell 133}
\tablewidth{0pt}
\tablehead{
\colhead{Date} & \colhead{{\it VLA} Configuration}   & \colhead{Frequency}   &
\colhead{Bandwidth} &
\colhead{Duration} &
\colhead{rms} &
\colhead{Obs.\ Code}\\
\colhead{} & \colhead{} & \colhead{(MHz)} & \colhead{(MHz)} & \colhead{(hours)} & \colhead{(mJy/bm)} & \colhead{}}
\startdata
2002 Jun 9  & B & 328.5/321.6 & 6.25/6.25 & 2.3 & 0.88\tablenotemark{a} & AC647\\
2002 Dec 9  & C & 1364.9/1435.1 & 25/25 & 3.1 & 0.065\tablenotemark{a} & AM735\\
2003 Aug 16 & A & 73.8/328.5 & 1.5/6.25 & 3.2 & 65.9/0.86 & AC647\\

\enddata
\tablenotetext{a}{The rms listed is for the map made by combining both frequencies.}
\label{tab:radio}
\end{deluxetable}

\begin{deluxetable}{lcccccc}
\tablewidth{0pt}
\tablecaption{X-ray Spectral Fits \label{tab:xspec}}
\tablehead{
\colhead{Region}&
\colhead{$kT_1$}&
\colhead{$kT_2$}&
\colhead{$Z$}&
\colhead{$\Gamma$}&
\colhead{$N_{\rm H}$}&
\colhead{$\chi^2$/dof}\\
\colhead{}&
\colhead{(keV)}&
\colhead{(keV)}&
\colhead{($Z_\odot$)}&
\colhead{}&
\colhead{($\times 10^{20}$cm$^{-2}$)}&
\colhead{}
}
\startdata
R1&1.10$^{+0.18}_{-0.08}$&2.21$^{+0.28}_{-0.21}$&1.27$^{+0.26}_{-0.21}$&...&5.1$^{+1.7}_{-1.3}$&154/127=1.21\\
R2&1.04$^{+0.27}_{-0.13}$&2.32$^{+0.15}_{-0.11}$&1.47$^{+0.26}_{-0.24}$&...&1.7$^{+1.5}_{-1.5}$&159/108=1.48\\
R3&1.06$^{+0.17}_{-0.07}$&2.21$^{+0.28}_{-0.21}$&1.41$^{+0.63}_{-0.42}$&...&2.5$^{+2.9}_{-2.4}$&90/88=1.02\\
R3&1.34$^{+0.10}_{-0.03}$&...&0.42$^{+1.62}_{-0.07}$&0.4$^{+2.2}_{-1.7}$&5.1$^{+2.5}_{-2.8}$&96/88=1.09\\
R4&2.82$^{+0.11}_{-0.16}$&...&1.27$^{+0.16}_{-0.15}$&...&3.7$^{+1.1}_{-1.1}$&157/126=1.25\\
R4&2.14$^{+0.23}_{-0.08}$&54.40$^{+\infty}_{-48.40}$&0.86$^{+0.27}_{-0.14}$&...&4.8$^{+1.3}_{-1.4}$&142/123=1.16\\
R4&2.23$^{+0.20}_{-0.14}$&...&0.76$^{+0.17}_{-0.14}$&0.27$^{+0.82}_{-0.89}$&5.6$^{+1.6}_{-1.5}$&141/123=1.14\\
R5&1.10$^{+0.28}_{-0.09}$&2.39$^{+0.96}_{-0.37}$&1.73$^{+0.99}_{-0.66}$&...&5.1$^{+4.1}_{-3.1}$&64/55=1.16\\
R6&1.11$^{+0.21}_{-0.08}$&2.22$^{+0.32}_{-0.21}$&0.91$^{+0.32}_{-0.24}$&...&(5.1)&62/72=0.81\\
Total\tablenotemark{a}&2.10$^{+0.03}_{-0.04}$&5.43$^{+0.25}_{-0.28}$&0.84$^{+0.09}_{-0.07}$/0.67$^{+0.06}_{-0.06}$\tablenotemark{b}&...&2.9$^{+0.3}_{-0.3}$&791/777=1.02\\
\enddata
\tablenotetext{a}{Fit to the total diffuse emission within the central
  1.9\arcmin\ (121~kpc).}
\tablenotetext{b}{Separate abundances given for lower and higher
  temperature components, respectively.}
\end{deluxetable}

\begin{deluxetable}{lccccc}
\tablewidth{0pt}
\tablecaption{Spectra Across the Wing Feature \label{tab:wings}}
\tablehead{
\colhead{Region}&
\colhead{$kT_1$}&
\colhead{$kT_2$}&
\colhead{$Z$}&
\colhead{$N_{\rm H}$}&
\colhead{$\chi^2$/dof}\\
\colhead{}&
\colhead{(keV)}&
\colhead{(keV)}&
\colhead{($Z_\odot$)}&
\colhead{($\times 10^{20}$cm$^{-2}$)}&
\colhead{}
}
\startdata
ND&3.00$^{+0.11}_{-0.11}$&...&1.10$^{+0.18}_{-0.16}$&(2.9)&98/89=1.10\\
ND&1.48$^{+0.61}_{-0.37}$&3.11$^{+0.12}_{-0.13}$&1.06$^{+0.20}_{-0.17}$&(2.9)&93/86=1.08\\
NU&3.02$^{+0.11}_{-0.12}$&...&1.43$^{+0.24}_{-0.21}$&(2.9)&86/76=1.19\\
NU&2.55$^{+0.11}_{-0.12}$&14.17$^{+\infty}_{-8.64}$&1.26$^{+0.28}_{-0.25}$&(2.9)&83/69=1.21\\
WD&3.69$^{+0.17}_{-0.18}$&...&1.35$^{+0.28}_{-0.24}$&(2.9)&51/66=0.77\\
WD&2.01$^{+0.76}_{-0.59}$&4.95$^{+12.73}_{-1.10}$&1.12$^{+0.32}_{-0.30}$&(2.9)&44/63=0.70\\
WU&3.75$^{+0.37}_{-0.27}$&...&0.60$^{+0.37}_{-0.27}$&(2.9)&935/30=1.17\\
WU&2.07$^{+0.68}_{-0.40}$&21.72$^{+\infty}_{-13.92}$&0.31$^{+0.25}_{-0.14}$&(2.9)&26/27=0.98\\
\enddata
\end{deluxetable}

\begin{deluxetable}{lccccc}
\tablewidth{0pt}
\tablecaption{Spectral Fits to Peripheral Regions\tablenotemark{a} \label{tab:acisi_spec}}
\tablehead{
\colhead{Region\tablenotemark{b}}&
\colhead{$kT_1$}&
\colhead{$kT_2$}&
\colhead{$Z$}&
\colhead{$\Gamma$}&
\colhead{$\chi^2$/dof}\\
\colhead{}&
\colhead{(keV)}&
\colhead{(keV)}&
\colhead{($Z_\odot$)}&
\colhead{}&
\colhead{}
}
\startdata
N. Clump&2.31$^{+0.37}_{-0.30}$&...&0.14$^{+0.15}_{-0.09}$&...&69/83=0.83\\
Hot Spot&6.74$^{+2.76}_{-1.24}$&...&0.66$^{+0.83}_{-0.65}$&...&64/59=1.09\\
Hot Spot&2.31$^{+1.06}_{-0.61}$&$35^{+\infty}_{-24}$&(0.66)&...&57/57=1.01\\
Hot Spot&2.72$^{+0.067}_{-0.79}$&...&(0.66)&1.19$^{+0.23}_{-0.38}$&58/57=1.02\\
NW Corner&4.11$^{+0.12}_{-0.12}$&...&0.46$^{+0.08}_{-0.07}$&...&477/433=1.10\\
SE Corner&4.59$^{+0.15}_{-0.15}$&...&0.34$^{+0.07}_{-0.07}$&...&445/419=1.06\\
\enddata
\tablenotetext{a}{Absorption fit to the best-fitting value to the total
  emission of $N_H = 2.9 \times 10^{20}$~cm$^{-2}$.}
\tablenotetext{b}{Regions labeled in Figure~\ref{fig:acisi_tmap}.}
\end{deluxetable}

\begin{deluxetable}{lccccc}
\tablewidth{0pt}
\tablecaption{Fluxes and Integrated Spectral Indices of Individual Features \label{tab:sindex}}
\tablehead{
\colhead{Feature}&
\colhead{$F_{330,{\rm A}}$\tablenotemark{a}\tablenotemark{b}}&
\colhead{$F_{1400,{\rm B}}$\tablenotemark{a}\tablenotemark{b}}&
\colhead{$\alpha_{330,{\rm A}-1400,{\rm B}}$\tablenotemark{c}}&
\colhead{$\alpha_{330,{\rm B}-1400,{\rm C}}$\tablenotemark{c}}&
\colhead{$\alpha_{74,{\rm A}-330,{\rm B}}$\tablenotemark{c}}\\
\colhead{}&
\colhead{(mJy)}&
\colhead{(mJy)}&
\colhead{}&
\colhead{}&
\colhead{}
}
\startdata
A133 Core&146.46 $\pm \, 2.53$&22.68 $\pm \, 0.19$&-1.29 $\pm \, 0.06$&...&...\\
A133 Core Jet\tablenotemark{d}&289.92 $\pm \, 3.31$&...&...&...&...\\
Relic/N. Lobe\tablenotemark{e}&3267.20 $\pm \, 7.27$&136.75 $\pm \, 0.16$&-2.20 $\pm \, 0.05$&-2.11 $\pm \, 0.05$&-1.60 $\pm \, 0.10$\\
S. Core\tablenotemark{f}&129.72 $\pm \, 3.06$&23.21 $\pm \, 0.16$&-1.19 $\pm \, 0.06$&-1.12 $\pm \, 0.05$\tablenotemark{f}&-2.10 $\pm \, 0.14$\tablenotemark{f}\\
S. Core N. Jet\tablenotemark{f}&91.9 $\pm \, 3.62$&...&-1.48 $\pm \, 0.16$&-2.49 $\pm \, 0.06$\tablenotemark{g}&...\\
S. Core S. Jet\tablenotemark{f}&60.9 $\pm \, 3.03$&...&...&-2.95 $\pm \, 0.07$\tablenotemark{g}&...\\
Arc&39.15 $\pm \, 2.87$&...&...&-1.96 $\pm \, 0.07$\tablenotemark{g}&...\\
S. Lobe&203.72 $\pm \, 8.02$&...&...&...&-1.65 $\pm \, 0.14$\\
W. Core&11.33 $\pm \, 1.75$&4.92 $\pm \, 0.15$&-0.58 $\pm \, 0.26$&...&...\\
\enddata
\tablenotetext{a}{Errors do not include calibration errors.}
\tablenotetext{b}{Subscripts indicate frequency in MHz and {\it VLA}
  configuration.}
\tablenotetext{c}{Subscripts indicate frequencies and {\it VLA}
  configurations used to calculate index.}
\tablenotetext{d}{Emission connecting the Abell~133 core to the relic,
  not detected at higher frequencies.}
\tablenotetext{e}{Region includes the Abell~133 core jet but excludes
  the core itself.}
\tablenotetext{f}{Region is difficult to isolate, so measurements are
  likely contaminated by nearby emission.}
\tablenotetext{g}{Measurements may suffer from missing 1.4~GHz flux,
  leading to artificially steep spectral indices.}
\end{deluxetable}

\clearpage

\clearpage

\end{document}